\documentclass[longauth]{aa} 
\usepackage{graphicx}
\usepackage{txfonts}
\usepackage[switch]{lineno}
\begin{document}
\title{First broadband characterization and redshift determination of the VHE blazar MAGIC J2001+439}
\author{
J.~Aleksi\'c\inst{1} \and
S.~Ansoldi\inst{2} \and
L.~A.~Antonelli\inst{3} \and
P.~Antoranz\inst{4} \and
A.~Babic\inst{5} \and
P.~Bangale\inst{6} \and
U.~Barres de Almeida\inst{6} \and
J.~A.~Barrio\inst{7} \and
J.~Becerra Gonz\'alez\inst{8,}\inst{25,}\inst{*} \and
W.~Bednarek\inst{9}
% ...
}
\institute { IFAE, Campus UAB, E-08193 Bellaterra, Spain
\and Universit\`a di Udine, and INFN Trieste, I-33100 Udine, Italy
\and INAF National Institute for Astrophysics, I-00136 Rome, Italy
\and Universit\`a  di Siena, and INFN Pisa, I-53100 Siena, Italy
\and Croatian MAGIC Consortium, Rudjer Boskovic Institute, University of Rijeka and University of Split, HR-10000 Zagreb, Croatia
\and Max-Planck-Institut f\"ur Physik, D-80805 M\"unchen, Germany
\and Universidad Complutense, E-28040 Madrid, Spain
\and Inst. de Astrof\'isica de Canarias, E-38200 La Laguna, Tenerife, Spain
\and University of \L\'od\'z, PL-90236 Lodz, Poland
\and Deutsches Elektronen-Synchrotron (DESY), D-15738 Zeuthen, Germany
\and ETH Zurich, CH-8093 Zurich, Switzerland
\and Universit\"at W\"urzburg, D-97074 W\"urzburg, Germany
\and Centro de Investigaciones Energ\'eticas, Medioambientales y Tecnol\'ogicas, E-28040 Madrid, Spain
\and Technische Universit\"at Dortmund, D-44221 Dortmund, Germany
\and Inst. de Astrof\'isica de Andaluc\'ia (CSIC), E-18080 Granada, Spain
\and Universit\`a di Padova and INFN, I-35131 Padova, Italy
\and Universit\`a dell'Insubria, Como, I-22100 Como, Italy
\and Unitat de F\'isica de les Radiacions, Departament de F\'isica, and CERES-IEEC, Universitat Aut\`onoma de Barcelona, E-08193 Bellaterra, Spain
\and Institut de Ci\`encies de l'Espai (IEEC-CSIC), E-08193 Bellaterra, Spain
\and Japanese MAGIC Consortium, Division of Physics and Astronomy, Kyoto University, Japan
\and Finnish MAGIC Consortium, Tuorla Observatory, University of Turku and Department of Physics, University of Oulu, Finland
\and Inst. for Nucl. Research and Nucl. Energy, BG-1784 Sofia, Bulgaria
\and Universitat de Barcelona, ICC, IEEC-UB, E-08028 Barcelona, Spain
\and Universit\`a di Pisa, and INFN Pisa, I-56126 Pisa, Italy
\and now at: NASA Goddard Space Flight Center, Greenbelt, MD 20771, USA and Department of Physics and Department of Astronomy, University of Maryland, College Park, MD 20742, USA
% ...
}
  \date{Received ... ; accepted ...}
  \abstract
  {}
  {We aim to characterize the broadband emission from 2FGL
  J2001.1+4352, which has been associated with the unknown-redshift blazar MG4 J200112+4352.
  Based on its gamma-ray spectral properties, 
  it was identified as a potential very high energy (VHE; $E > 100$ GeV) gamma-ray emitter.
  We investigate whether this object is a VHE emitter,
  characterize its gamma-ray spectrum, and study the broadband emission within the one-zone synchrotron
  self-Compton (SSC) scenario, which is commonly used to describe the emission in blazars. 
  Moreover, we also intend to determine the redshift of this object, which is a crucial parameter for its scientific interpretation.}
  {The source was observed with MAGIC first in 2009 and later in 2010 within a multi-instrument observation campaign.
  The MAGIC observations yielded 14.8 hours of good quality stereoscopic data.
  Besides MAGIC, the campaign involved, observations with \emph{Fermi}-LAT, \emph{Swift}-XRT/UVOT, the optical telescopes KVA, Goddard Robotic Telescope, Galaxy View observatory, 
  Crimean Astrophysical observatory, St. Petersburg observatory, and the Owens Valley Radio Observatory. The object was monitored at radio, optical and gamma-ray energies during 
  the years 2010 and 2011. We characterize the radio to VHE spectral energy distribution and quantify the multiband variability and correlations over short (few days) and 
  long (many months) timescales. We also organized 
  deep imaging optical observations with the Nordic Optical Telescope in 2013 to determine the source redshift.}
  {The source, named MAGIC J2001+439, is detected for the first time at VHE with MAGIC at a
  statistical significance of 6.3 $\sigma$ ($E > 70$ GeV) during a 1.3-hour 
  long observation on 2010 July 16. 
  The multi-instrument observations show variability in all energy bands with the highest amplitude of variability in the X-ray and VHE bands. 
  Besides the variability on few-day timescales, the long-term monitoring of MAGIC J2001+439 shows that,
  the gamma-ray, optical, and radio emissions gradually decreased on few-month
  timescales from 2010 through 2011, 
  indicating that at least some of the radio, optical and gamma-ray emission is produced
  in a single region by the same population of particles. 
  We also determine for the first time the redshift of this BL Lac object
  through the measurement of its host galaxy during low blazar activity.
  Using the observational evidence that the luminosities
  of BL Lac host galaxies are confined to a relatively narrow range,
  we obtain $z = 0.18 \pm 0.04$.
  Additionally, we use the \emph{Fermi}-LAT and MAGIC gamma-ray spectra to provide
  an independent redshift estimation,
  $z = 0.17 \pm 0.10$.
  Using the former (more accurate) redshift value,
  we adequately describe the broadband emission with a one-zone SSC model for different activity states
  and interpret the few-day timescale variability as produced by changes in the high-energy component of the electron energy distribution.}
  {}
  \maketitle
\section{Introduction}

    Blazars are radio-loud active galactic nuclei (AGN) with relativistic jets pointing 
    towards the observer \citep[see e.g.][]{urry1995}.
    They are the most common extragalactic sources detected in the very high energy 
    (VHE; $E >$ 100 GeV) gamma-ray range.
    The spectral energy distributions (SEDs) of blazars show a double-bump shape.
    The first bump peaks at optical/X-ray frequencies and is attributed to synchrotron radiation from
    relativistic electrons. On the other hand, the origin of the second bump, which peaks at gamma-ray
    energies, is still under debate. 
    Leptonic models are generally favored. In these models, 
    the high-energy (HE; $E >$ 100 MeV) radiation is produced by inverse Compton (IC) of primary HE electrons scattering 
    off low-energy photons. The origin of target low-energy photons may be synchrotron radiation of the primary electrons 
    themselves in the synchrotron self-Compton (SSC) scenario \citep{band1985, maraschi1992, bloom1996}, 
    or seed photons produced outside of the jet in the external Compton (EC) scenario \citep{dermer1993, sikora1994}.
    However, hadron-driven emission is also possible \citep[e.g.][]{mannheim1992, mucke2003}.
    The emission would include proton, muon, and pion synchrotron radiation, as well as production of gamma rays from
    neutral pion decays, electrons, and positrons generated in charged pion decays that result from 
    photon-hadron collisions. The hadronic models require total jet powers that are typically about 1-2 orders of
    magnitude higher than for the leptonic models.

    Blazars are separated into two categories by the equivalent widths of their optical 
    emission lines, BL Lac objects and flat-spectrum radio quasars (FSRQs).
    The BL Lac objects show featureless optical spectra with weak or no emission lines that are possibly 
    masked by a strong non-thermal emission from the relativistic
    jet, while the FSRQs display prominent broad emission lines in their optical spectra. 
    The absence of emission/absorption lines
    makes it very difficult to determine the redshift for distant BL Lac objects, 
    which often precludes detailed studies on the nature,
    intrinsic characteristics of individual objects, and substantially hampers and/or 
    biases blazar population/unification studies.

    The flux of the VHE gamma-ray photons coming from a distant source is attenuated by 
    electron-positron pair creation due to interaction with the extragalactic background light 
    \citep[EBL;][]{gould1966, stecker1969, fazio1970, hauser2001}.
    The EBL is the sum of the stellar and dust emission integrated over cosmic time.
    The EBL photon density carries information about the cosmic history of the star formation 
    rate on galaxy evolution.
    Several EBL models have been proposed in the past few years 
    \citep{stecker2006, franceschini2008, gilmore2009, kneiske2010, finke2010, dominguez2011}.
    The VHE gamma-ray absorption is energy dependent and increases strongly with redshift.
    Therefore, the observed VHE spectra from distant sources are distorted with respect to the
    intrinsic source spectra.
    The distances of unknown redshift BL Lac objects can be estimated by comparing the GeV and TeV spectra 
    and assuming a specific EBL model \citep{elisa2011}.
    This estimation is based on the measurement of the intrinsic source spectrum with the \emph{Fermi} Large Area Telescope (LAT)
    at energies below $\sim$10-30 GeV, where there is little or no EBL absorption.

    There are only $\sim$50 blazars significantly detected at VHE\footnote[1]{\url{http://tevcat.uchicago.edu}}.
    This very low number of known VHE blazars is a consequence of the difficulty of performing
    sensitive scans over large portions of the sky with 
    Imaging Atmospheric Cherenkov Telescopes (IACTs), which have narrow
    field of view (3$^\circ$--5$^\circ$) cameras and only $\sim$1000
    hours of moonless time per year with good weather conditions.
    On the other hand, more than 1000 HE gamma-ray emitting blazars have been detected
    with \emph{Fermi}-LAT \citep{abdo2009b, nolan2012}, and many of them have been identified
    (based on their spectral properties) as potential VHE emitting sources \citep{FHL}.

    The object 0FGL J2001.0+4352 was initially one of the unidentified \emph{Fermi}-LAT sources included in
    the \emph{Fermi} bright source list \citep{abdo2009b}.
    This source was first detected only above 1 GeV with a photon flux
    $F_{1\rm{GeV}} =$ ($7.8 \pm 1.2$) $\times 10^{-9}$ ph cm$^{-2}$ s$^{-1}$
    between 1 and 100 GeV. Early on, this source was identified by the \emph{Fermi}-LAT
    collaboration as a source expected to exhibit VHE
    emission, which is information that was shared with the H.E.S.S., MAGIC and VERITAS collaborations in
    2009 October. 
    This information triggered observations with MAGIC, which led to the first
    VHE detection of this source in 2010 July \citep[see][]{mariotti, karsten2010, karsten2011}.
    This source was initially
    designated MAGIC J2001+435, although
    we change its name to MAGIC J2001+439 in this paper to properly follow the
    IAU guidelines for naming astronomical objects.

    The latest \emph{Fermi}-LAT catalogs confirmed the brightness and hardness of the gamma-ray 
    spectra of this source. 
    In the second \emph{Fermi}-LAT source catalog \citep[2FGL,][]{nolan2012}, this source is denoted 2FGL J2001.1+4352,
    and its spectrum is characterized with a 
    power-law  function $\mathrm{d}N/\mathrm{d}E \propto E^{- \Gamma}$ with 
    $\Gamma = 1.90 \pm 0.03$ above 100 MeV.
    This source is also present in the first \emph{Fermi} 
    HE LAT catalog
    \citep[1FHL,][]{FHL},
    where it is denoted 1FHL J2001.1+4353, and the
    spectrum is characterized by a power-law function with $\Gamma = 2.38 \pm 0.18$ above 10 GeV,
    extending to VHE with a flux above 100 GeV of
    $F_{100\rm{GeV}} =$ ($2.2^{+ 1.9}_{- 1.2}$) $\times 10^{-11}$ ph cm$^{-2}$ s$^{-1}$ \citep{FHL}. 
    \cite{bassani2009} found a counterpart consistent with 
    the radio bright source MG4 J200112+4352\footnote[2]{MG4 J200112+4352 is located at RA(J2000) = 20h 01m 12.9s, Dec(J2000) = +43d 52m 53s 
    (NASA/IPAC Extragalactic Database, \url{http://ned.ipac.caltech.edu}).}
    from the NRAO VLA Sky Survey \citep[NVSS;][]{condon1998} $20$ cm wavelength image.
    The object MG4 J200112+4352 is only 0.01 deg away from the location of 2FGL J2001.1+4352
    (which has a 95$\%$ confidence level position uncertainty of 0.02 deg) and is consistent with
    the \emph{Swift} and the XMM Slew positions in the X-ray band.
    The source was identified as a BL Lac object using spectroscopic 
    observations with the 1.52 m optical telescope from the Bologna Astronomical observatory
    and was classified as a high-frequency-peaked BL Lac object (HBL) by \cite{bassani2009}.
    The source redshift remained undetermined due to the low signal-to-noise ratio of their observations.
    Yet, they found indications of a slope change in the optical spectrum, which could be 
    interpreted as a  non-thermal component merging with the light from the host galaxy. 
    Based on this feature, they gave a rough estimate of $\sim$0.2 for the redshift of this source. 
    More recently, \cite{shaw} used higher quality optical observations to derive a lower limit 
    for the redshift of $z$ $>$ 0.11, based on the non-detection of the host galaxy, which was 
    assumed to be a giant elliptical galaxy with an absolute R-band magnitude 
    of $M_{R} = - 22.5 \pm 0.5$.

    \begin{figure*}[htpb]
    \begin{tabular}{ccc}
    \begin{minipage}[h]{0.333\textwidth}
      \includegraphics[width=\textwidth]{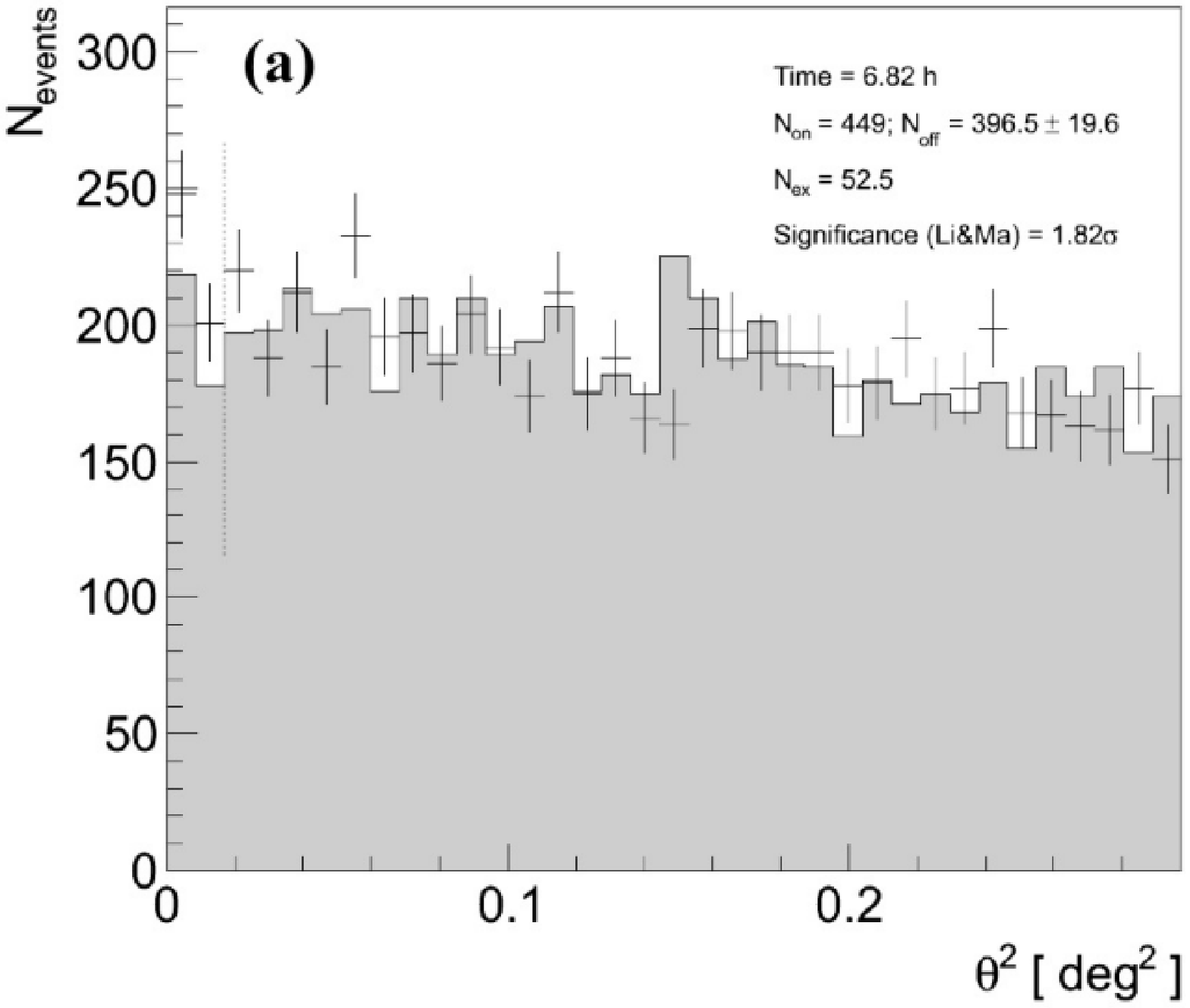}
      \label{fig1}
    \end{minipage}
    \begin{minipage}[h]{0.333\textwidth}
      \includegraphics[width=\textwidth]{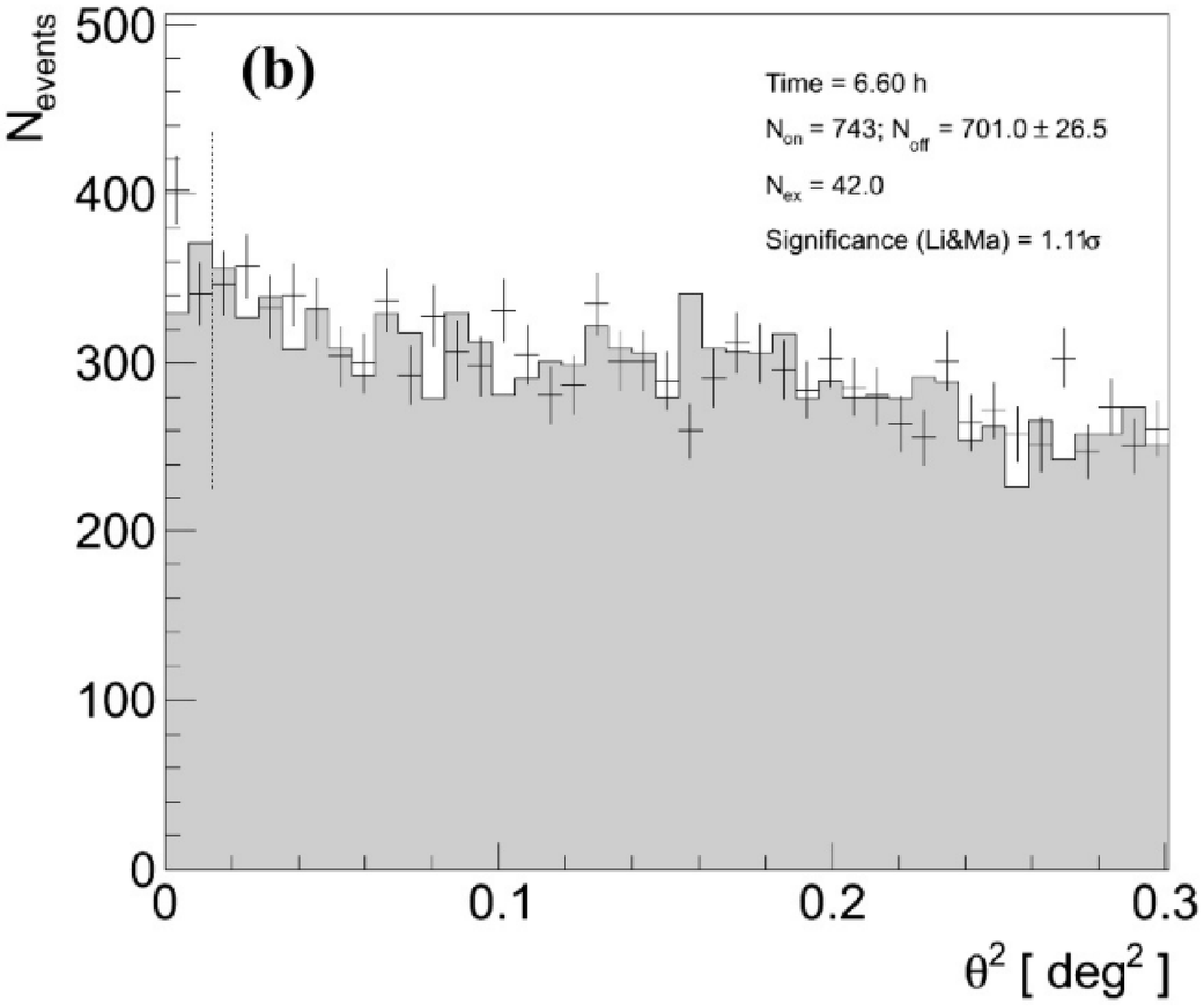}
      \label{fig2}
    \end{minipage}
    \begin{minipage}[h]{0.333\textwidth}
      \includegraphics[width=\textwidth]{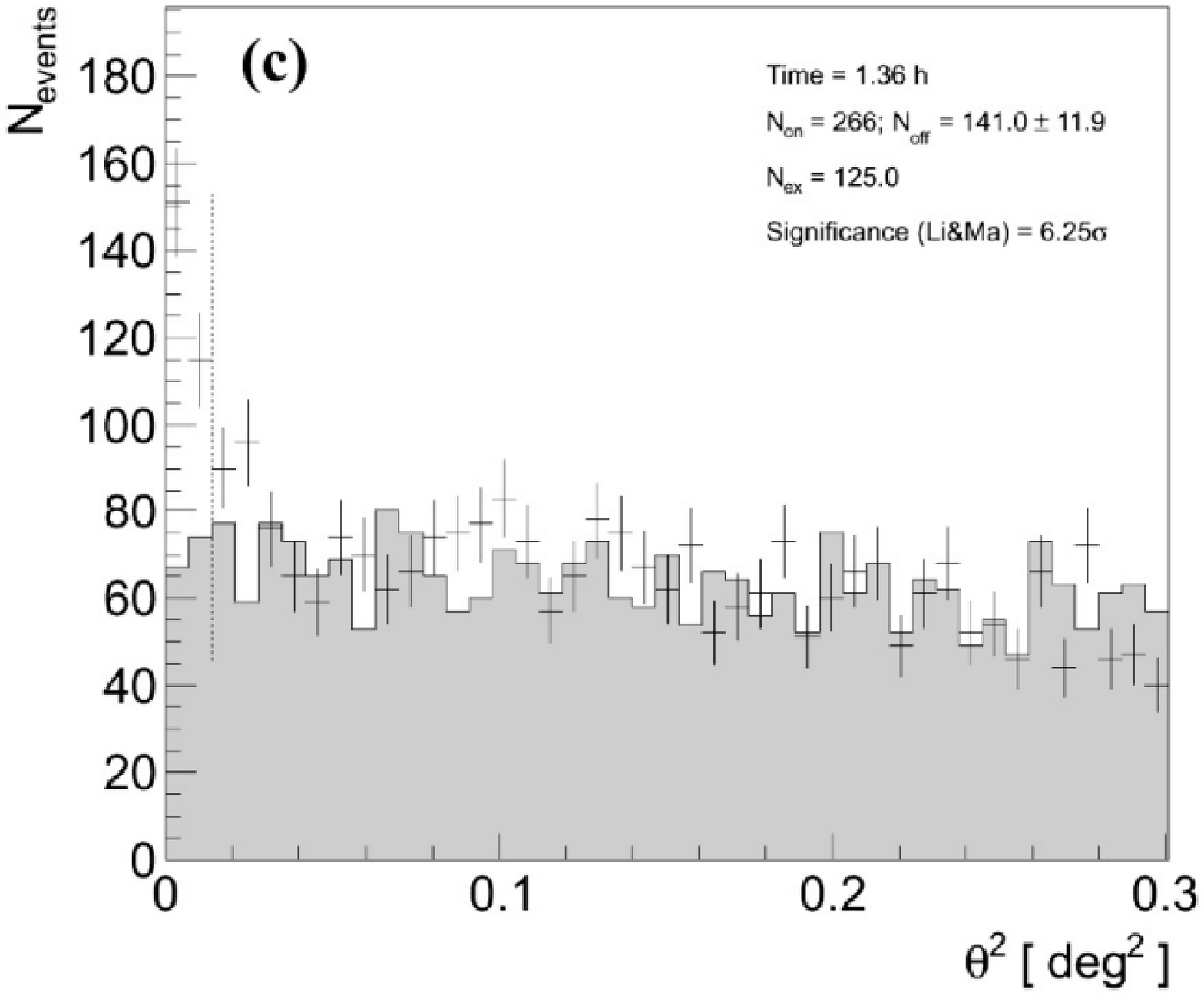}
      \label{fig1}
    \end{minipage}
    \end{tabular}
    \caption{Theta-squared distributions of MAGIC J2001+439 observed on 2009 November (panel (a), energy threshold is $E$ $>$ 100 GeV)
    between July and September excluding 2010 July 16 (panel (b), $E$ $>$ 70 GeV) and on 2010 July 16 (panel (c), $E$ $>$ 70 GeV).
    Crosses represent the event distribution from the source, while the gray histogram the measured background.
    The signal region is indicated by the vertical dotted line.}
    \end{figure*}

    In this paper, we report the results from a multi-wavelength
    (MWL) campaign from summer 2010, providing coverage from the radio up
    to the VHE gamma-ray band and leading to the first VHE detection of this source.
    The multi-instrument observations allowed us to characterize, 
    the radio to VHE broadband SED of this object for the first time. 
    We also report on the 
    multiband variability and correlation properties during this campaign and follow-up 
    observations performed during the years 2010 and 2011. Moreover, we report the
    first measurement of the redshift for this source through the
    detection of its host galaxy with the 2.5m Nordic Optical Telescope
    (NOT) during low blazar activity. Additionally, we also estimate
    the redshift of this object  using the HE and VHE gamma-ray spectra, as measured by \emph{Fermi}-LAT and MAGIC.
    We then use the measured SED and 
    redshift information to characterize the radio to VHE broadband emission within a standard 
    one-zone SSC scenario and investigate the origin of the detected variability.
    In this paper, we assume cosmological parameters $H_{0} = 67$ km s$^{-1}$ Mpc$^{-1}$,
    $\Omega_{m} = 0.315$, $\Omega_{\Lambda} = 0.685$ \citep{planck2013}.


\section{MAGIC observations and results}
\subsection{Observation and data analysis}

    The MAGIC stereoscopic system consists of two IACTs with a mirror dish diameter of 17 m
    located at the Roque de los Muchachos, La Palma in Canary Island (28.8\textordmasculine N, 17.8\textordmasculine W at 2200 m a.s.l.).
    The MAGIC telescopes have been operating in stereoscopic mode since autumn 2009, which provided 
    integral sensitivity of 0.76$\%$ of the Crab Nebula flux above 300 GeV for 50 hour observation time \citep{aleksic2012a}.

    The object MAGIC J2001+439 was observed between 2009 November 7 and 26 for a total of 9.0 hours.
    The MAGIC observations were also performed in a MWL campaign between 2010 July 6 and September 8 for a total of 14.4 hours.
    The data were taken with zenith angles in the range 20 deg -- 40 deg
    in 2009 November and with zenith angles in the range 15 deg -- 30 deg
    during the campaign in 2010 July -- September.
    The observations were carried out in wobble mode \citep{fomin1994},
    where the target source position has an offset of 0.4$\textordmasculine$ from the camera center.
    The direction of the wobble offset between two symmetric sky locations is alternated every 20 minutes to minimize systematic 
    errors originating from possible exposure inhomogeneities.

    The data were analyzed using the standard analysis chain \citep{aleksic2012a} with the MAGIC Analysis and Reconstruction Software \citep[MARS;][]{moralejo2009, roberta2013}.
    Camera images were cleaned using a sum image-cleaning method \citep{saverio2011, roberta2011}.
    This algorithm originated from the concept of the sum trigger \citep{rissi2009, haefner2011}.
    In this procedure, the signals are clipped in amplitude and all possible combinations of 2, 3 and 4 neighboring pixels 
    in the camera are summed up.
    If the sum of the charges is above a certain threshold within a short time interval, these pixels are considered to belong to the shower image.
    The clipping ensures that afterpulses or strong night sky background fluctuations do not dominate the summed pixels.
    Generally, the sum image-cleaning method recovers more pixels than the standard method.
    This is important for reconstructing shower images of low-energy gamma rays.

\subsection{Results}

    Figure \ref{fig1} shows the distribution of the squared angular distance ($\theta^{2}$) between
    the reconstructed arrival directions of the events and the real source position in the camera.

    We found an excess of events $N_{ex}$ = 125.0 $\pm$ 20.2 in the energy range above 70 GeV
    in the observation on 2010 July 16 in which the effective observation time was 1.36 hours (see Figure 1 (c)).
    This gamma-ray excess yields a signal significance of 6.3 $\sigma$ calculated using Eq.17 of \cite{li1983}.
    When correcting for the seven observations (trials) performed in the
    MWL campaign, we find a post-trial signal significance of 6.0$\sigma$, 
    hence implying the first detection of VHE gamma rays from 2FGL J2001.1+4352.
    The time-averaged integral photon flux above 200 GeV corresponds to $\sim$9$\%$ of the Crab Nebula flux.
    The detected position of the excess (RA(J2000): 20.021 $\pm$ 0.001 h, Dec(J2000): 43.879 $\pm$ 0.010\textordmasculine) is consistent with 
    the position of 2FGL J2001.1+4352\footnote[3]{2FGL J2001.1+4352 is
    located at RA(J2000) = 20.019 h, Dec(J2000)= 43.879\textordmasculine in the 2FGL
    catalog \citep{nolan2012}.} within 0.02\textordmasculine.
    The distribution of the gamma-ray excess is consistent with a point-like source.
    The source was not detected during the rest of the observing campaign (see Figure 1 (b)).
    In the data between 2010 July and September (excluding 2010 July
    16), the significance of the excess in 8.0 hours of observations is
    1.1 $\sigma$ above the energy threshold of 70 GeV.  Including the
    observations from  2010 July 16, the significance (above 70 GeV) of the accumulated dataset is 4 $\sigma$.
    The data collected in 2009 November led to 6.8 hours of effective observation time, where 
    we measure a gamma-ray excess above the energy threshold of 100 GeV at a significance level of 1.8 $\sigma$ (see Figure 1 (a)).
    The slightly higher energy threshold in the 2009 MAGIC
    observations with respect to that of the 2010 observations is due
    to the different zenith angle range for these two sets of
    observations.

    The differential spectrum from the flare on 2010 July 16 can be described by a simple power law:
    \begin{eqnarray}
    \frac{\mathrm{d}N}{\mathrm{d}E} = f_{0} \times \left(\frac{E}{200\: {\rm GeV}}\right)^{- \Gamma}
    \end{eqnarray}
    with flux normalization $f_{0}$ = $(1.9 \pm 0.4) \times 10^{-10}$ cm$^{-2}$ s$^{-1}$ TeV$^{-1}$
    and photon index $\Gamma$ = $2.8 \pm 0.4$.
    The spectrum is fitted by a power-law function between 78 and 500 GeV.
    The systematic uncertainties in the spectral measurements with MAGIC stereo observations are 11$\%$ in the 
    normalization factor (at 300 GeV) and 0.15-0.20 in the photon index.
    The error on the flux does not include uncertainty on the energy scale.
    The energy scale of the MAGIC telescopes is determined with a precision of about 17$\%$ at low energies ($E < 100$ GeV)
    and 15$\%$ at medium energies ($E > 300$ GeV).
    Further details are reported in \cite{aleksic2012a}.
    We corrected for our limited energy resolution and energy bias using the Tikhonov unfolding algorithm \citep{albert2007}.
    The result is shown in Figure \ref{fig2}.
    
    \begin{figure}[htpb]
    \centering
    \includegraphics[width=\hsize]{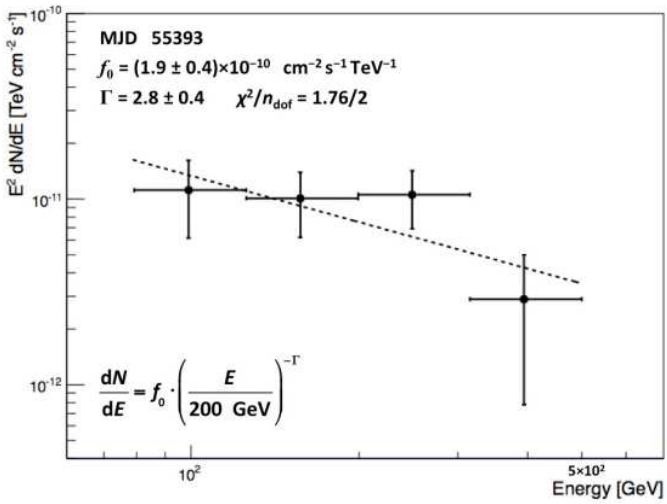}
    \caption{VHE differential energy spectrum of MAGIC J2001+439
      observed on 2010 July 16 with the MAGIC stereo system. The
      parameter values from the power-law fit are reported in the
      legend. }
    \label{fig2}
    \end{figure}

\section{Multiband variability and correlations}

\subsection{Instrumentation and data analysis}
\subsubsection{\emph{Fermi}-LAT}

    The \emph{Fermi}-LAT is a pair conversion telescope designed to cover the energy band from 20 MeV to values greater than 300 GeV \citep{atwood2009},
    which operates in survey mode.
    The data were analyzed with the \emph{Fermi} Science Tools package version v9r27p1 available from the \emph{Fermi} Science Support Center 
    (FSSC)\footnote[4]{\url{http://fermi.gsfc.nasa.gov/ssc}}.
    For this analysis, only events belonging to the Pass7 Source class and located in a circular region of interest of
    10$\textordmasculine$ radius of 2FGL J2001.1+4352 were selected.
    Moreover, 
    events with zenith angles greater than 100$\textordmasculine$ were removed
    to reduce the contamination from the Earth-limb gamma-rays,
    which are produced by cosmic rays interacting with the upper atmosphere, 
    and time intervals during which the rocking angle of the spacecraft exceeded 52$\textordmasculine$ were excluded.
    The background model used to extract the gamma-ray signal includes a Galactic diffuse emission component and isotropic
    components (including residual cosmic rays), which were modeled using the files
    gal\_2yearp7v6\_v0.fits and isotropic iso\_p7v6source.txt that are publicly available\footnote[5]{\url{http://fermi.gsfc.nasa.gov/ssc/data/access/lat/BackgroundModels.html}}.
    The normalizations of the components comprising the total background model were allowed to vary freely during the spectral point fitting.
    The spectral fluxes were derived with the post-launch instruments response functions P7\_V6\_SOURCE and by applying
    an unbinned maximum likelihood technique \citep{mattox1996} to events in the energy range spanning 300 MeV to 300 GeV.
    All the sources from the 2FGL catalog located within 10$\textordmasculine$ radius were included in the model of the region.
    The source position and initial spectrum parameters 
    in the XML file were set to those of the 2FGL catalog.
    Flux upper limits at 95$\%$ confidence level were computed for those time intervals with a test statistic 
    \citep[TS;][]{mattox1996} value below four.
    The systematic uncertainty in the flux is dominated by the systematic uncertainty in the effective area,
    which is estimated as 10$\%$ at 100 MeV, 
    is increased to 5$\%$ at 560 MeV, and is increased to 10$\%$ at 10 GeV \citep{ackermann2012}.
    The systematic uncertainties are smaller than the statistical uncertainties of the data points in the light curve and spectra.

    \begin{figure*}[htpb]
    \centering
    \includegraphics[width=13.0cm]{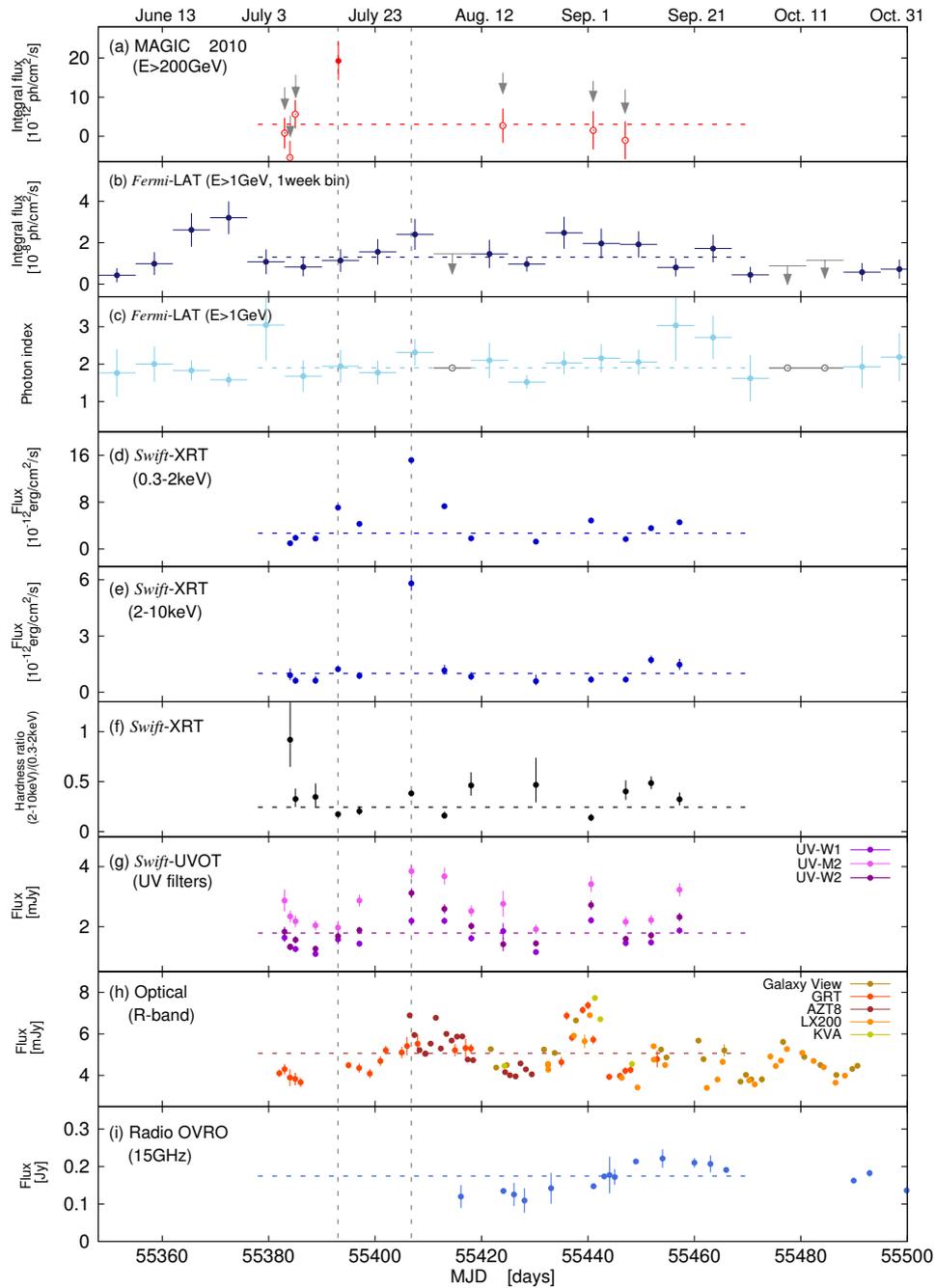}
    \caption{Multiband light curves of MAGIC J2001+439 during the observing campaign in 2010.
    All light curves show fluxes from single night observations, except for \emph{Fermi}-LAT.
    (a) MAGIC light curve above 200 GeV. The red filled circle depicts
    the flux during the VHE flare on 2010 July 16 (the only
    significant detection),
    while the open circles correspond to flux points with excess significances between $-1.3$ and $1.6$ $\sigma$ \citep[calculated according to][Eq.17]{li1983}.
    The gray arrows report the 95$\%$ confidence level upper limits,
    calculated using a photon index of 2.8.
    (b) \emph{Fermi}-LAT light curve above 1 GeV with a weekly binning.
    The gray arrows report the flux upper limits at 95$\%$ confidence
    level, which were calculated (using the photon index of 1.9,
    reported for this source in
    the 2FGL catalog) for the
    time intervals with TS$<$4.
    (c) \emph{Fermi}-LAT photon index computed with a weekly
    binning. The gray open circles denote the assumed photon index for
    the calculation of the upper limits.
    (d) \emph{Swift}-XRT light curve in the energy range from 0.3 to 2 keV.
    (e) \emph{Swift}-XRT light curve in the energy range from 2 to 10 keV.
    (f) Hardness ratio (2-10 keV) / (0.3-2 keV).
    (g) \emph{Swift}-UVOT light curves for the three UV filters.
    (h) Optical R-band light curves from different telescopes (see legend).
    (i) Radio light curve at 15 GHz from the OVRO telescope.
    The horizontal dotted lines show the result of a fit with a constant function (the UV-$W2$ flux points were used for the \emph{Swift}-UVOT light curve).
    The first and second gray vertical dashed lines denote July 16 and July 29 respectively.}
    \label{fig3}
    \end{figure*}

    \begin{figure*}[htpb]
    \centering
    \includegraphics[width=14cm]{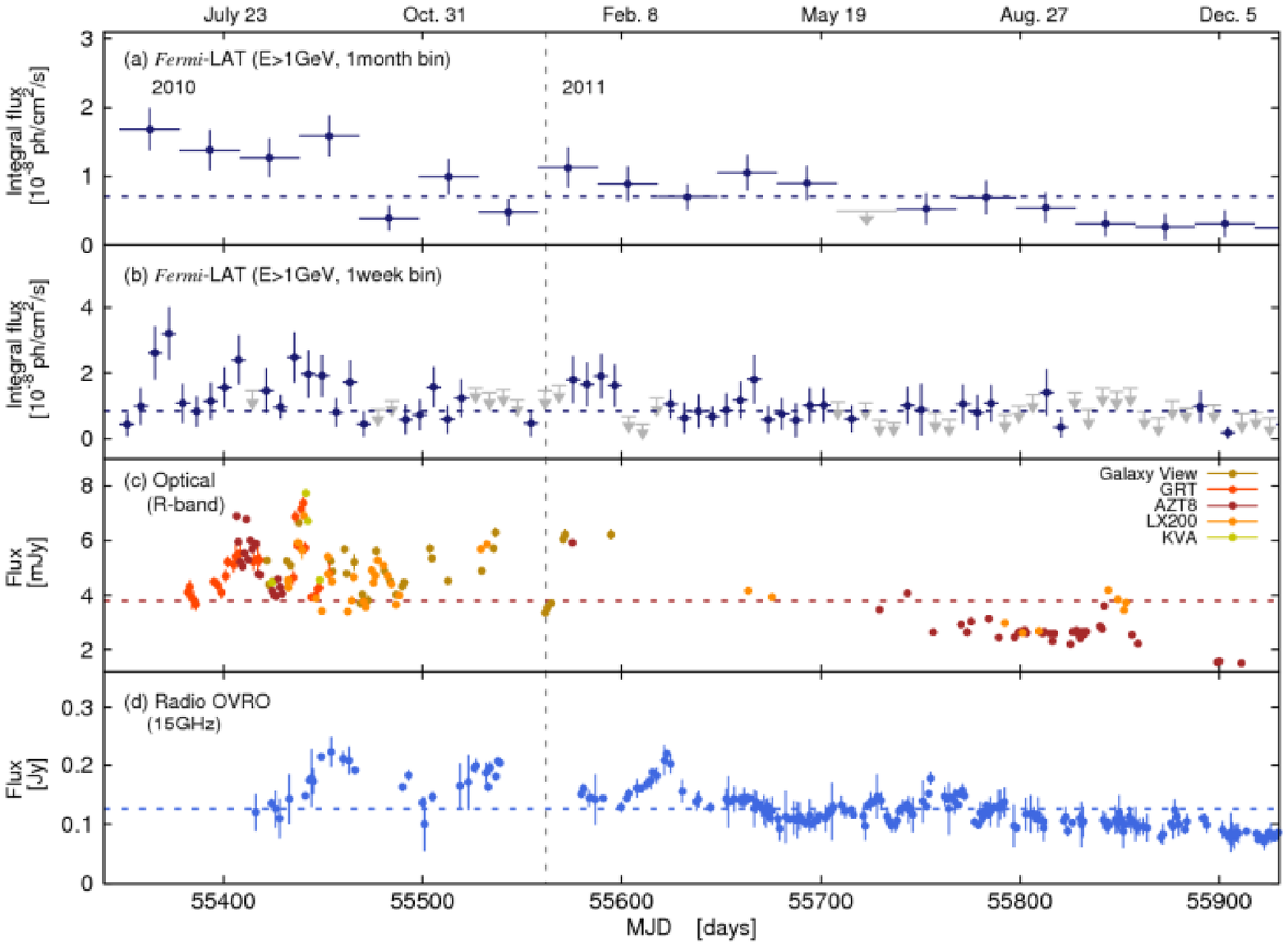}
    \caption{Long-term light curves of MAGIC J2001+439 in 2010 and 2011.
    (a) \emph{Fermi}-LAT light curve above 1 GeV with a monthly binning.
    (b) \emph{Fermi}-LAT light curve above 1 GeV with a weekly binning.
    The gray arrows report the flux upper limits at 95$\%$ confidence
    level, which were calculated for the time intervals with TS$<$4.
    (c) Optical R-band light curves.
    (d) Radio light curve at 15 GHz.
    The horizontal dotted lines show the result of a fit with a constant function.
    See caption of Figure 3 for further details.}
    \label{fig4}
    \end{figure*}

\subsubsection{\emph{Swift}}

    \emph{Swift} is equipped with three telescopes: the Burst Alert Telescope \citep[BAT;][]{barthelmy2005}, which covers the 15 -- 150 keV range,
    the X-ray telescope \citep[XRT;][]{burrows2005}, which covers the 0.3 -- 10 keV energy range, and the Ultra-Violet/Optical telescope \citep[UVOT;][]{roming2005}, 
    which covering the 180 -- 600 nm wavelength range with six bandpass filters.

    The results reported here relate to measurements performed with XRT and UVOT. The BAT instrument is not sufficiently sensitive to
    detect this object: MAGIC J2001+439 is in neither the 70-month BAT catalog \citep{2013ApJS..207...19B} nor the BAT transient monitor paper \citep{2013ApJS..209...14K}.

    The \emph{Swift} satellite observed the source 15 times in 2010. All XRT observations
    were carried out using the Photon Counting (PC) readout mode. The dataset 
    was first processed with the XRTDAS software package (v.2.9.3)
    developed at the ASI Science Data Center (ASDC) and
    distributed by HEASARC within the HEASoft package (v. 6.15.1).
    Event files were calibrated and cleaned with standard
    filtering criteria with the {\it xrtpipeline} task using the 
    calibration files available in the \emph{Swift} CALDB version \emph{20140120}.
    The average spectra were extracted from the cleaned event
    files. Events for the spectral analysis were selected within a circle
    of 20 pixel ($\sim$46 $\arcsec$) radius, which encloses about 90$\%$ of the 
    point-spread function (PSF),
    centered on the source position. The background was extracted from a 
    nearby circular region of 40 pixel radius. Two observations were excluded
    from the analysis due the very short exposure leading to
    insufficient number of counts.
    The ancillary response files (ARFs) were generated with the {\it
    xrtmkarf} task applying corrections for PSF losses and CCD defects
    using the cumulative exposure map. 
    Before the spectral fitting, the 0.3 -- 10 keV source energy spectra were binned to 
    ensure a minimum of 20 counts per bin.
    The results from the \emph{Swift}-XRT observations during the MWL campaign in 2010
    are summarized in Table A1. The X-ray count rates and hardness ratios
    for the intra-night light curve from July 16 (the day with the
    VHE flare) were  extracted from the automatic \emph{Swift}
    XRT analysis for \emph{Fermi}-LAT
    sources\footnote[6]{\url{http://www.swift.psu.edu/monitoring}}.

%
     
   The \emph{Swift}-UVOT observations on MG4 J200112+4352 were
   conducted with UV filters only (namely $W1$, $M2$, and $W2$).
   We performed an aperture photometry analysis for all filters in all the observations using the standard UVOT
   software distributed within the HEAsoft 6.13 package and the
   calibration included in the version \emph{20140120} of the \emph{Swift} CALDB.
   Counts were first extracted from an aperture of 5\arcsec\, radius for all filters and converted to fluxes using the standard zero points \citep{poole08}.
   We excluded a nearby contaminating star
   from the aperture \citep[USNO B1.0 1338-0359172, photographic magnitudes B2=15.81, R2=14.98; ][]{UB1},
   which is the only one within the extraction region in our UV images.
   We also tried the ``curve of growth'' method, which is included in the official software for apertures with FWHM radii for each filter.
   We obtained compatible results, except for larger errors using the second method.
   The fluxes were then de-reddened using $E(B-V)$\,=\,0.498 \citep{schlegel1998, schlafly2011} and 
   with $A_{\lambda}/E(B-V)$ ratios calculated for UVOT filters using the mean galactic interstellar extinction curve from \citep{Fitzpatrick1999}. 
   The results were also carefully checked for other possible contaminations.
   No intra-observation variability has been detected taking the nearby star into account.
   The results of \emph{Swift}-UVOT are summarized in Table A2.
    The results obtained for the individual observations show a peak in the source's broadband SED
    at the M2 frequency with the energy flux for the M2 filter being
    up to a factor of two larger than for the W1
    filter. The energy flux for M2 is up to 50\% higher than that for
    W2.
    Comparable ``narrow features'' in the SED, which are not expected in 
    regular synchrotron bumps from leptonic theoretical scenarios 
    (see Figures \ref{fig13}, \ref{fig14}, and \ref{fig15})
    are observed in the \emph{Swift}-UVOT results for other
    sources \citep[e.g. see Figure 4 in][]{PKS1424VERITASPaper}, hence
    we conclude that it is an instrumental effect (not related to the
    object we are studying). This effect could be due to the source
    having a B$-$V that is out of the validity range indicated by
    \citet{poole08} for their flux calibrations in the UV bands.
    This instrumental effect, however, does not have any impact on the main results reported here.


\subsubsection{Optical band}

    The optical R-band flux density was monitored during the campaign by several instruments.
    These optical aperture photometry observations
    were performed with the 35 cm optical telescope at the KVA observatory on La Palma
    (that operates in close collaboration with the MAGIC telescopes), 
    the Goddard Robotic Telescope (GRT) at the Goddard Geophysical and
    Astronomical observatory, the
    Galaxy View observatory, a 70 cm AZT-8 telescope at the Crimean Astrophysical observatory, and a 40 cm LX-200 telescope in St. Petersburg observatory.
    All the observed R-band magnitudes were corrected with the
    Galactic extinction of $A_{\lambda} = 1.219$ mag
    \citep{schlafly2011}, but the optical data points shown in Figure
    \ref{fig3} are not host-galaxy corrected.
    The typical statistical error is $\sim$0.03 mag, which is comparable to the systematic error related to these measurements.
    For the typical fluxes measured during these observations (15.7
    mag), the relative statistical error of the flux measurements is $\sim$3\%.

    As reported in section 3.3.2, the contribution from the host galaxy is small in comparison to the blazar emission.
    The overall host galaxy emission, when corrected for the Galactic extinction, would be $\sim$0.5 mJy.
    However, the contribution of the host galaxy to the measured blazar emission would depend on the details of the optical
    observations and data reduction.
    For the particular case of KVA (which used a fixed aperture radius of 4 arcsec for these observations),
    the host galaxy contribution to the measured blazar fluxes would be $\sim$0.3 mJy.
    Such estimates for the other telescopes were not performed, but one would expect similar values within $\sim$0.1 mJy. 
    Therefore, the subtraction of the constant emission from the host galaxy would only shift the fluxes down by $\sim$6\%
    with an additional difference among instruments at the level of $\sim$2\%.
    We considered these small offsets not essential for the results reported in this paper.

\subsubsection{Radio band}


    Radio monitoring at 15~GHz was performed with the 40~m telescope at
    the Owens Valley Radio Observatory (OVRO) through an ongoing blazar
    monitoring program\footnote[7]{\url{http://www.astro.caltech.edu/ovroblazars}}.
    Observations of MAGIC J2001+439 commenced on 2010 August 8
    and were scheduled approximately twice per week.
    The observation and calibration procedures are described in detail in \cite{richards2011}.
    Flux densities were measured in a 3~GHz wide band centered at 15.0~GHz, 
    using off-axis dual beam optics with azimuth double switching to remove 
    atmospheric and ground interference. 
    The flux density scale is determined from regular observations of 3C~286 
    by assuming the value of 3.44~Jy at 15.0 GHz \citep{baars1977}, 
    which leads to about 5\% scale uncertainty and is not included in our error bars.
    Individual uncertainties are estimated from an error model that accounts 
    for non-thermal random errors in addition to the measured scatter during the observation. 
    The error model works well on average but occasionally produces excessively
    conservative uncertainties, particularly during poor weather. 
    About 2\% of the radio data was excluded due to bad weather.
%

\subsection{Results}
\subsubsection{Overall multiband light curves}

    This MWL campaign was conducted over 2.5 months in 2010.
    The overall MWL light curves of MAGIC J2001+439 during the campaign are shown in Figure \ref{fig3}.
    From all the MAGIC observations, only the one from 2010 July 16 yielded a significant detection. 
    The other observations yielded excesses with a signal significance below 2 $\sigma$.
    In the light curve, we show the photon fluxes and also the  95$\%$ confidence
    level flux upper limits calculated night by night during 2010 July to September.
    The upper limits were derived by assuming a power-law spectrum with a photon index of $\Gamma = 2.8$, 
    which is the one measured for the observation on 2010 July 16. 
    We note that the computed upper limits depend on the assumed photon index.
    When using a photon index of $\Gamma = 4.0$, the upper limits increase $\sim$5\%. 
    The integral flux above 200 GeV on 2010 July 16 is 
    $F_{200 \rm{GeV}} = (1.9 \pm 0.5) \times 10^{-11}$ ph cm$^{-2}$ s$^{-1}$, which
    corresponds to $\sim$9\% the flux of the Crab Nebula.
    The integral flux upper limit during the non-detection period 
    in 2009 November 
    shows $F_{200 \rm{GeV}} < 1.0 \times 10^{-11}$ ph cm$^{-2}$ s$^{-1}$ above 200 GeV.
    The \emph{Fermi}-LAT light curve is plotted with a temporal bin width of one week
    in the energy range from 1 to 300 GeV.
    The one-week averaged \emph{Fermi}-LAT photon index (for a power-law function
    above 1 GeV) was also computed (see third panel in Figure \ref{fig3}),
    indicating the absence of spectral variability (on weekly
    timescales) during the three-month observing campaign.

    The multiband light curves from Figure \ref{fig3} show that 
    MAGIC J2001+439 is variable at all energy bands 
    with the largest flux variations in the X-ray light curve.
    It is worth noting the 2-10 keV X-ray flux during the VHE
    flare from July 16 is only 
    twice as large as that measured
    during previous \emph{Swift}-XRT observations in 2010 July, while the 2-10
    keV flux from July 29 (MJD 55406) is $\sim$five times higher than that of July 16. 
    The 0.3-2 keV X-ray light curve shows the same trend as the variation of the 2-10 keV X-ray light curve, 
    and the hardness ratio also changes in this campaign.
    Unfortunately, bad weather conditions precluded MAGIC observations on July 29.
    Moreover, the OVRO monitoring observations started soon after the announcement 
    of the first VHE detection \citep{mariotti}.

    In Figure \ref{fig4}, we report the long-term light curves at GeV gamma-rays,
    optical, and radio bands, as a result of a dedicated optical/radio follow-up
    during more than one year. The emission in these three energy bands shows
    a gradual decrease from 2010 through 2011. Further details on
    the long-term variability and correlations are discussed in
    sections \ref{Fvar} and \ref{DCF}.

\subsubsection{July 16 intranight light curves}

\begin{figure}[htpb]
    \centering
    \includegraphics[width=\hsize]{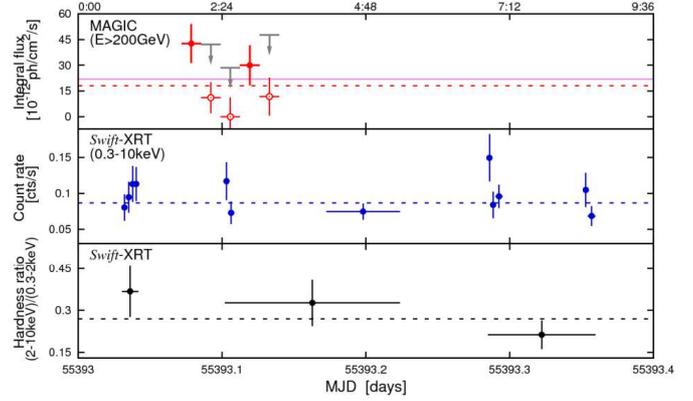}
    \caption{Intra-night multiband light curves of MAGIC J2001+439 observations on 2010 July 16.
    Top panel: VHE gamma-ray flux above 200 GeV, as measured by MAGIC.
    The open circles depict the fluxes with excess significances between 0 and 1.2 $\sigma$ \citep[calculated according to][Eq.17]{li1983}.
    The gray arrows report the 95$\%$ confidence level upper limits.
    The magenta solid line depicts 10$\%$ of the Crab Nebula flux, while the red dashed line
    reports the result of a fit with a constant function ($\chi^{2}$/$n_{\rm{dof}}$ = 9.3/4).
    Middle panel: X-ray count rate in the energy range 0.3-10 keV, as measured by
    \emph{Swift} XRT. The blue dashed line depicts the result of a fit with a
    constant function ($\chi^{2}$/$n_{\rm{dof}}$ = 12.3/11). Bottom panel: Hardness ratio (2-10 keV)/(0.3-2 keV). The black
    dashed line depicts the result of a fit with a constant
    function ($\chi^{2}$/$n_{\rm{dof}}$ = 2.9/2). The X-ray count rate and hardness ratios were extracted from the automatic \emph{Swift}
    XRT analysis for LAT sources$^{6}$}
    \label{fig5}
    \end{figure}

    The intranight light curves of MAGIC J2001+439 on 2010 July 16 are shown in Figure \ref{fig5}.
    The top panel shows the light curve of the MAGIC observations with an interval of 20 minutes in the energy range above 200 GeV.
    A constant fit to the data yielded a flux above 200 GeV of $F_{200 \rm{GeV}} = (1.8 \pm 0.5) \times 10^{-11}$ ph cm$^{-2}$ s$^{-1}$ with a
    $\chi^{2}$/$n_{\rm{dof}}$ = 9.3/4, which corresponds to a $\chi^{2}$ probability of $P_{\chi}$ = 5.4$\%$. 
    This indicates that the MAGIC light curve is consistent with a constant flux hypothesis within 95$\%$ confidence level.
    The \emph{Swift}-XRT intranight light curve and the hardness ratio plots are shown in the same figure, respectively.
    The fit with a constant to the X-ray count rate gives a $\chi^{2}$ of 12.3 for $n_{\rm{dof}}$ = 11 ($P_{\chi}$ = 34$\%$), and to
    the hardness ratio curve gives a $\chi^{2}$ of 2.9 for $n_{\rm{dof}}$ = 2 ($P_{\chi}$ = 23$\%$).
    We did not detect any statistically significant intra-night variability.

\subsubsection{Fractional variability}
\label{Fvar}

    To quantify the energy dependence of variability amplitudes, we computed the fractional variability amplitude
    of the light curves for each spectral band \citep{vaughan2003}.
    The fractional variability amplitude $F_{var}$ is calculated as
    \begin{eqnarray}
    F_{var} = \sqrt{\frac{S^{2} - e^{2}}{F^{2}}},
    \end{eqnarray}
    where $S^{2}$ is the total variance of the light curve, $e^{2}$ is the mean squared error, $F$ is the mean flux.\\

    The uncertainty of $F_{var}$ is defined as
    \begin{eqnarray}
    \Delta F_{var} = \sqrt{F_{var}^{2} + err(\sigma_{NXS}^{2})} - F_{var},
    \end{eqnarray}
    as reported in Poutanen et al. (2008).\\
    The $err(\sigma_{NXS}^{2})$, which is the error in the normalised excess (NXS) variance, is given by equation 11 of Vaughan et al. (2003)
    \begin{eqnarray}
    err(\sigma_{NXS}^{2}) = \sqrt{\left( \sqrt{\frac{2}{N}} \frac{e^{2}}{F^{2}} \right)^{2} + \left(\sqrt{\frac{e^{2}}{N}} \frac{2F_{var}}{F} \right)^{2}}.
    \end{eqnarray}

    This methodology to quantify the variability has some caveats. The
    fractional variability $F_{var}$ is determined for the temporal
    bin and the source sampling from the light curves used, and
    the source variability might actually depend on that. In other
    words, a densely sampled light curve with very small temporal bins
    might allow us to see flux variations that are hidden otherwise,
    and hence we might obtain a larger $F_{var}$. In the set of light
    curves shown in Figure \ref{fig3} and Figure \ref{fig4}, one can
    see that some energy bands are better sampled than others. In
    particular, the information from the gamma-ray band is limited with only
    seven observations performed with MAGIC and fluxes on weekly/monthly time intervals
    (instead of daily time intervals) obtained with \emph{Fermi}-LAT.  Another
    caveat is that the fractional variability given by equation (2)
    expects data points with similar error bars (within one dataset). 
    A few data points
    with substantially (by factors of a few) larger error bars would
    have a larger impact in $e^{2}$ than in $S^{2}$, hence 
    biasing $F_{var}$ towards lower values. In our multi-instrument dataset,
    the band that is most affected by this effect is the 15 GHz radio
    light curve 
    provided by OVRO. Despite the above-mentioned caveats, the
    $F_{var}$ from \cite{vaughan2003} is a useful methodology
    to quantify in a simple way the variability in the different
    energy bands sampled during this observing MWL campaign.

\begin{figure}[htpb]
    \centering
    \includegraphics[width=\hsize]{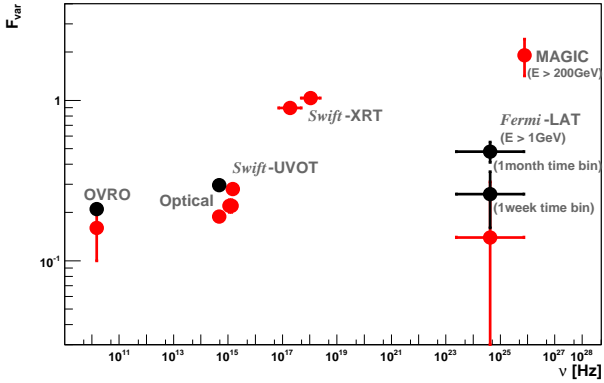}
    \caption{Fractional variability vs. energy. 
    The red data points refer to the MWL campaign light curves
    in 2010 July to September
    reported in Figure \ref{fig3}, while the black data points refer to
    the long-term light curves in 2010 to 2011 reported in Figure \ref{fig4}.}
    \label{fig6}
    \end{figure}

    Figure \ref{fig6} shows the fractional variability obtained with the
    data reported in the light curves from Figure \ref{fig3} and
    Figure \ref{fig4}. The obtained $F_{var}$ values for all energy bands are listed in Table 1.
    During the MWL campaign in summer 2010,
    we measured significant $F_{var}$ values for the optical, UV, X-ray, and VHE gamma-ray bands with
    the variability being greatest at X-ray and VHE.
    The $F_{var}$ at VHE is dominated by the large flux increase during the flaring activity on
    July 16, although this $F_{var}$ value needs to be taken with caveats
    due to the small number of observations in
    comparison with those performed at other energy bands and the lack of VHE observations during the strong
    X-ray flare on July 29.
    As for the long-term behavior, we measure significant $F_{var}$ values 
    in the three bands sampled, namely radio, optical and
    HE gamma-rays. The fractional variability obtained for these three bands is 
    similar.

    \begin{table}[htpb]
    \caption{Fractional variability amplitudes for different energy
      bands and two time periods.}
    \centering
    \begin{tabular}{l c}
    \hline\hline
    Energy bands & $F_{var} \pm \Delta F_{var}$ \\
    \hline
    (campaign in 2010) & \\
    MAGIC ($E$ > 200 GeV) & 1.92 $\pm$ 0.50 \\
    \emph{Fermi}-LAT one-week bins ($E$ > 1 GeV) & 0.14 $\pm$ 0.17 \\
    \emph{Swift}-XRT (2 -- 10 keV) & 1.03 $\pm$ 0.05 \\
    \emph{Swift}-XRT (0.3 -- 2 keV) & 0.90 $\pm$ 0.01 \\
    \emph{Swift}-UVOT (W2) & 0.28 $\pm$ 0.02 \\
    \emph{Swift}-UVOT (M2) & 0.22 $\pm$ 0.02 \\
    \emph{Swift}-UVOT (W1) & 0.22 $\pm$ 0.02 \\
    Optical (R-band) &  0.19 $\pm$ 0.01 \\
    Radio (15 GHz) & 0.16 $\pm$ 0.06 \\
    \hline
    (long-term in 2010 -- 2011) & \\
    \emph{Fermi}-LAT one-month bins ($E$ > 1 GeV) & 0.48 $\pm$ 0.07 \\
    \emph{Fermi}-LAT one-week bins ($E$ > 1 GeV) & 0.26 $\pm$ 0.10 \\
    Optical (R-band) & 0.30 $\pm$ 0.01 \\
    Radio (15 GHz) & 0.21 $\pm$ 0.01 \\
    \hline\hline
    \\
    \end{tabular}
    \label{table:1}
    \end{table}


\subsubsection{Multiband correlations}
\label{DCF}

    We quantified the correlation among the MWL light curves shown in Figure \ref{fig3} and Figure \ref{fig4} by applying the
    Discrete Correlation Function (DCF) technique from
    \cite{edelson1988} to investigate the correlation
    between different energy bands for different time lags.
    For each of these pairs, we can compute the unbinned discrete correlation functions (UDCF),
    \begin{eqnarray}
    UDCF_{ij} = \frac{(F_{a_{i}} - F_{a})(F_{b_{j}} - F_{b})}{\sqrt{(S_{a}^{2} - e_{a}^{2})(S_{b}^{2} - e_{b}^{2})}},
    \end{eqnarray}
    where $F_{a_{i}}$ and $F_{b_{j}}$ are the data pair in the bin associated with the pairwise time lag $\Delta t_{ij} = t_{j} - t_{i}$,
    $F_{a}$ and $F_{b}$ are the mean flux values, $S_{a}$ and $S_{b}$ are the standard deviations, and 
    $e^{2}_{a}$ and $e^{2}_{b}$ are the mean measurement errors squared, respectively.\\
    The DCF for a given time lag of $\tau$ is then constructed as
    \begin{eqnarray}
    DCF(\tau) = \frac{1}{M} \sum_{ij} UDCF_{ij} (\Delta t_{ij}),
    \end{eqnarray}
    where the sum runs over the $M$ pairs of observations separated by $\tau - \Delta \tau / 2 \leq \Delta t_{ij} \leq \tau + \Delta \tau / 2$,
    where $\Delta \tau$ is the chosen bin width.\\ 
    The uncertainty on the value of the DCF in a given bin is calculated as the RMS variance of all the contributing $UDCF_{ij}$ about the value $DCF (\tau)$,
    \begin{eqnarray}
    \sigma_{DCF} (\tau) = \frac{1}{M-1} \left( \sum \left[ UDCF_{ij} - DCF (\tau) \right]^{2} \right)^{\frac{1}{2}}.
    \end{eqnarray}\\

    \begin{figure*}[htpb]
    \centering
    \includegraphics[width=11.0cm]{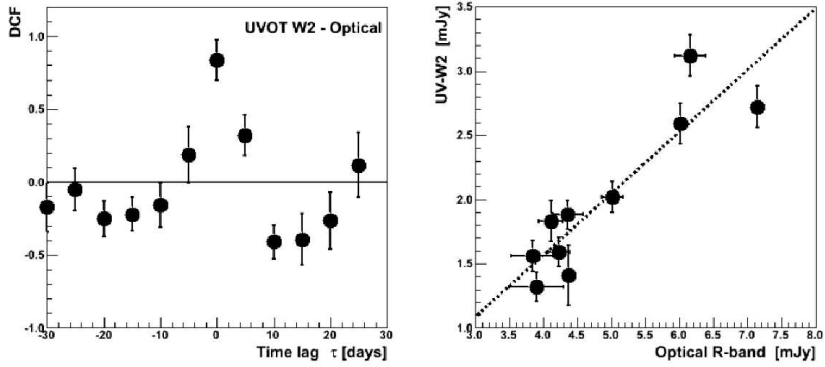}
    \caption{Discrete correlation function for different time lags computed for the UV (W2) and optical (R) data obtained during the MWL campaign in 2010 and
    the flux linear correlation plot for a time lag of zero. 
    The dotted line shows the best fit with a linear function.}
    \label{fig7}
    \end{figure*}

    \begin{figure*}[htpb]
    \centering
    \includegraphics[width=17.1cm]{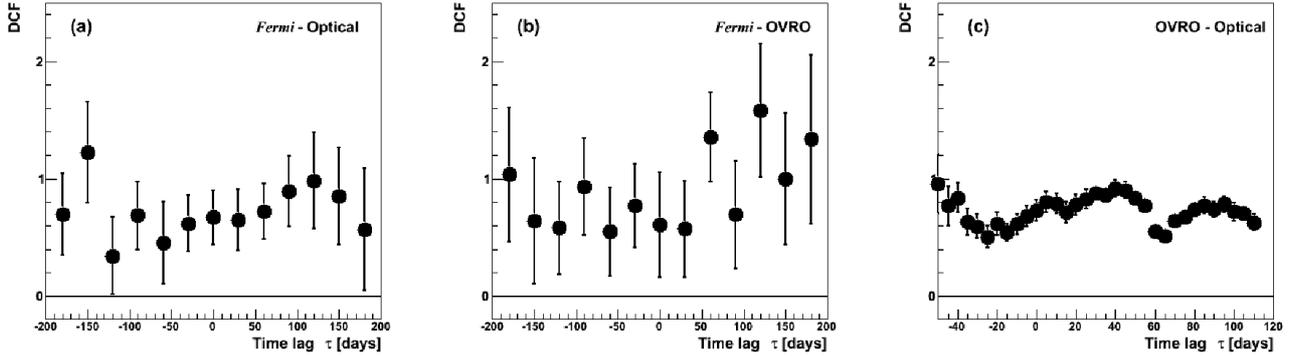}
    \caption{Discrete correlation function  for different time lags
    for the long-term light curves from Figure \ref{fig4}.
    (a) One-month-averaged \emph{Fermi}-LAT fluxes ($E$ > 1 GeV) vs. one-month-averaged optical R-band fluxes.
    (b) One-month-averaged \emph{Fermi}-LAT fluxes ($E$ > 1 GeV)
    vs. one-month-averaged radio 15 GHz fluxes.
    (c) Radio 15 GHz fluxes vs. optical R-band fluxes (only single
    observations occurring during the same day were used). }
    \label{fig8}
    \end{figure*}

    \begin{figure*}[htpb]
    \centering
    \includegraphics[width=17.1cm]{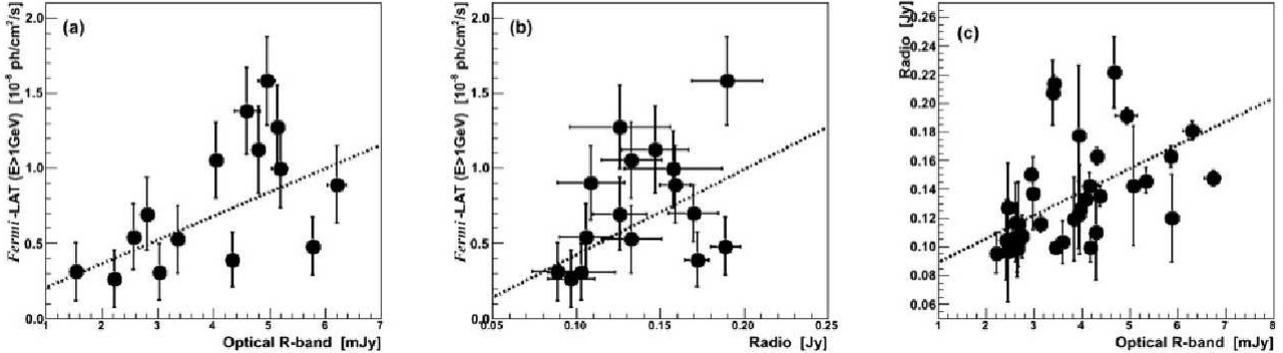}
    \caption{Flux-flux linear correlation plots for a time lag of zero
    for the long-term light curves reported in Figure \ref{fig4}.
    (a) One-month-averaged \emph{Fermi}-LAT fluxes ($E$ > 1 GeV) vs. one-month-averaged optical R-band fluxes.
    (b) One-month-averaged \emph{Fermi}-LAT fluxes ($E$ > 1 GeV) vs.
    one-month-averaged radio 15 GHz fluxes.
    (c) Radio 15 GHz fluxes vs. optical R-band fluxes. (Only single
    observations occurring during the same day were used.)
    The dotted line shows the best fit with a linear function.}
    \label{fig9}
    \end{figure*}

    We investigated the correlation among the different bands shown in
    Figure \ref{fig3}, obtaining a significant correlation only for the UV vs. optical band. 
    Figure \ref{fig7} shows 
    that the DCF for time lags of $\pm$30 days for the UV (W2) and optical (R) band.
    We used 5-day bins for the time lags, which minimizes the impact of the 2--3 day time
    gaps in the UV data.
    The plots show clearly that the significant correlation occurs
    only for a time lag zero. The Pearson's correlation coefficient\footnote[8]{This is a standard formulation that one can find 
    in section 14.5 of \cite{press1992}.}
    for these two datasets (depicted in the right panel in Figure \ref{fig7})
    is $r = 0.90$ with an accidental probability $P$ of
    no-correlation$^{8}$ of 0.04\% (with $n_{\rm{dof}} = 8$).
    The significant correlation is not surprising,
    given the proximity of these two energy bands.

    The DCF plots for long-term light curves from Figure \ref{fig4} are
    reported in Figure \ref{fig8}. The three panels show an overall positive
    correlation for all time lags, which is produced by the clear
    long-term decrease in the radio/optical/gamma-ray fluxes over many
    month timescales. The DCF for \emph{Fermi} vs. radio and \emph{Fermi}
    vs. optical cover $\pm$0.5 years with 30-day time lag bins. The large
    error bars in the DCF values, which are caused by the relatively large error bars in the
    gamma-ray fluxes, preclude the investigation of any temporal
    structure in the correlation.

    The DCF for the optical vs. radio flux cover time lags from -50
    days to +120 days with a temporal bin of 5 days.
    The asymmetry is driven by the result that the radio
    observations started about one month after the optical
    observations, and they extend further in time than the optical
    observations. This means that there is a lower tolerance to apply negative time shifts (i.e. the
    optical light curve is shifted to earlier times) and a higher tolerance to
    apply positive time shifts (i.e. the optical light curve is shifted to
    later times). Given the large
    number of data points and the relatively small single-night
    measurement errors, the errors in the DCF values are small, 
    which indicates a temporal structure on the top of the overall
    positive correlation. 
    The DCF is highest in the time lag range
    from 0 to +50 days with a maximum value at about +40 days.
    To investigate this DCF peak at about 40 days, we
    computed the normalized optical and radio long-term light curves,
    where the data flux values of each measurement are divided by the
    mean flux of the entire light curve.
    There are some structures in the optical and radio light curves 
    that are similar in amplitude and are better aligned when shifting 
    the optical light curve by +40 days;
    yet, there are also several other structures, which occur in one band 
    and not in the other. 
    Moreover, we note that +40 days is the minimum time lag
    needed to get the optical light curve starting at the same day as the radio light curve.
    Given the different length and density of observations for
    these two bands, we cannot make definite conclusions about the
    temporal structure observed in the DCF for these two bands, apart
    from the positive correlation produced by the long-term decrease
    in the light curves. Further studies would require more homogeneous 
    and better sampled light curves.

    Figure \ref{fig9} shows flux-flux linear correlation plots of time lag zero for long-term light curves.
    The derived Pearson's correlation coefficients are $r = 0.59$ for
    the GeV gamma-ray vs. optical R-band, $r = 0.43$ for the GeV gamma-ray vs. radio, and
    $r = 0.48$ for the radio vs. the optical R-band.
    The accidental probabilities are $P = 2.0\%$ (with number of
    degrees of freedom $n_{\rm{dof}} = 13$), $P = 10.0\%$ (with $n_{\rm{dof}} = 14$),
    and $P = 0.3\%$ (with $n_{\rm{dof}} = 35$) for GeV/optical, GeV/radio and radio/optical, respectively.
    These values indicate that there is a marginally significant
    positive linear correlation among the GeV/optical
    bands, and a much more significant positive correlation for the radio/optical bands.
    The higher significance for the correlation
    between the radio/optical bands is due to the larger number of
    measurements and smaller uncertainties in the measured fluxes (as implied by formulae 5, 6, and 7). 
    

\subsection{Redshift measurement of the blazar MAGIC J2001+439}

    Since the redshift of MAGIC J2001+439 was still uncertain, we used two independent methods to 
    determine it.

\subsubsection{Redshift estimation using the gamma-ray spectrum}

\begin{figure}[htpb]
    \centering
    \includegraphics[width=\hsize]{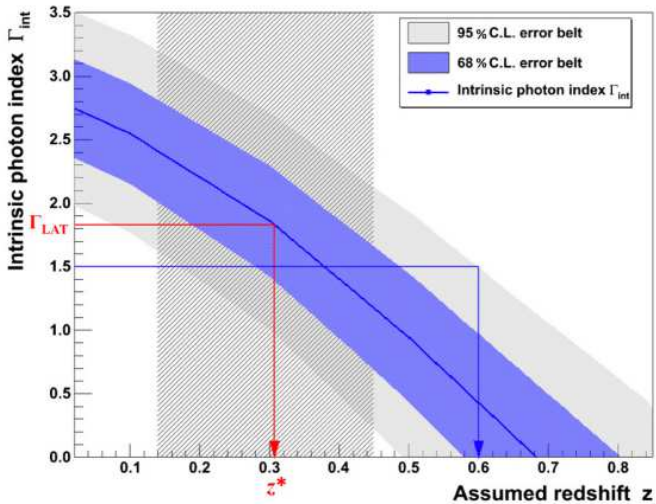}
    \caption{Intrinsic photon index $\Gamma_{\rm{int}}$ as a function
    of the redshift assuming the EBL model from
    \cite{franceschini2008}. The blue and gray filled areas
    correspond to the 1 $\sigma$ (68\%) and 2 $\sigma$ (95\%)
    confidence error belts, respectively.
    The red arrow indicates the estimated redshift $z^{*}$, which is
    determined by the redshift value at which $\Gamma_{\rm{int}} = \Gamma_{\rm{LAT}} = 1.83 \pm 0.18$.
    The diagonal shaded area shows the error range of $z^{*}$, where
    the error takes both the errors on the MAGIC VHE
    photon index and the \emph{Fermi}-LAT HE photon index into account.
    See text for calculation of the reconstructed redshift $z_{\rm {rec}}$ using the value $z^{*}$.
    The blue arrow indicates the 95$\%$ confidence level upper limit
    on the redshift, which is obtained when $\Gamma_{\rm{int}} = 1.5$,
    within the 2 $\sigma$ (95\%)
      confidence error belt.}
    \label{fig10}
    \end{figure}

    The redshift of a gamma-ray source can be estimated (or at least
    constrained) using the measured gamma-ray spectra, once a particular
    EBL model is assumed. In this work, we adopted the EBL model from  \cite{franceschini2008}.
    If we assume that the intrinsic source gamma-ray spectrum can be expressed by a simple 
    power-law function $\mathrm{d}N/\mathrm{d}E \propto E^{- \Gamma_{\rm{int}}}$,
    where the fitted intrinsic photon index is $\Gamma_{\rm{int}}$,
    one can set upper limits on the source redshift under the assumption that 
    the intrinsic source gamma-ray spectrum cannot be harder than 1.5.
    This limit is physically motivated from shock
    acceleration arguments, as discussed in \cite{aharonian2006}.
    As shown in Figure \ref{fig10}, when taking 
    the uncertainty in the measured VHE spectrum with MAGIC into account,
    we find that this assumption yields an upper limit on the redshift
    of $z$ $< 0.6$ with a 95$\%$ confidence level.

    Another estimate on the redshift can be obtained using the measured
    gamma-ray spectra with \emph{Fermi} and MAGIC, as reported in \cite{elisa2011}.
    We analyzed \emph{Fermi}-LAT data from MAGIC J2001+439 between 2010 July 1 and August 1 and obtained a
    one-month-averaged \emph{Fermi}-LAT spectrum.   This spectrum has
    a spectral index of $\Gamma_{\rm {LAT}} = 1.83 \pm 0.18$, when
    being characterized by a power-law function in the energy range
    between 300 MeV and 30 GeV. The integral flux is 
    $F_{300\rm{MeV}} =$ (3.9 $\pm$ 0.8) $\times 10^{-8}$ ph cm$^{-2}$ s$^{-1}$ in the energy range above 300 MeV.
    We define the redshift of $z^{*}$ by requiring that the power-law
    index of the observed VHE gamma-ray spectrum, when corrected for the EBL absorption, is 
    equal to the power-law index of $\Gamma = 1.83$ observed by
    \emph{Fermi}-LAT at energies which are not affected by the EBL
    absorption. 
    As shown in Figure \ref{fig10}, this procedure leads to $z^{*} = 0.31 \pm 0.16$. 
    In this calculation, we did not consider instrumental systematic errors in the determination of the spectral
    indices from MAGIC and \emph{Fermi}-LAT.
    We followed the prescription given in \cite{elisa2011} with the difference that, 
    we also considered the uncertainty in the
    power-law index from the HE spectrum measured with \emph{Fermi}, in
    addition to the uncertainty in the power-law index from the VHE
    spectrum when computing the uncertainty in the parameter $z^{*}$.
    Therefore, our uncertainty in the parameter  $z^{*}$ is larger, but more reliable than the one that would have
    been derived following \cite{elisa2011}.
    To determine the reconstructed redshift $z_{\rm {rec}}$,
    we used the empirical relation reported in \cite{elisa2011} that
    relates the true redshifts of known distance sources $z_{\rm {true}}$ with their $z^{*}$ values, $z^{*} = A + Bz_{\rm {true}}$, 
    where $A = 0.036 \pm 0.014$ and $B = 1.60 \pm 0.14$.
    Using this prescription, we obtain a reconstructed redshift
    $z_{\rm {rec}} = 0.17 \pm 0.10$.
    The systematic uncertainty related to this method is estimated to be 0.05 
    in the redshift value.

\subsubsection{Redshift determination using the measured flux of the host galaxy}
\label{RedshiftMeasurement2}

    The MAGIC J2001+439 (MG4 J200112+4352) was observed by the NOT telescope on
    2013 June 13 to study the host galaxy.
    To increase the detection probability, the observations were timed to coincide with an optical
    low state (R $\sim$ 16.8, corresponding to F $\sim$1.8 mJy) of the target.  We obtained nine images, each with
    900s exposure time, through the I-band filter using the ALFOSC
    instrument equipped with a 2048$\times$2048 E2V chip with a gain factor
    of 0.327 $e^-/$Analog to Digital Units (ADU),
    and readout noise of 4.2$e^-$. The total field of view 
    of ALFOSC is 6$\farcm$5$\times$6$\farcm$5, and the pixel
    scale is 0$\farcs$19/pixel. 
    The transparency of the atmosphere remained constant for all the 
    observations, which allowed for photometric measurements.
    The images were bias-subtracted and flat-fielded with
    twilight flats, after which the fringe pattern was removed using an
    archival fringe map. Individual images were then registered using 13 stars over the 
    field of view and co-added. The resulting image has a total
    exposure time of 2h 15min and FWHM of 0$\farcs$72.

    The calibration of the field was obtained from I-band observations
    of MAGIC J2001+439 and BL~Lac in photometric conditions with
    the 72'' Perkins Telescope at Lowell Observatory on 2013 November 2 and 3. 
    We first used the BL~Lac comparison star sequence in
    \cite{1996A&AS..116..403F} to calibrate five stars in the field of
    MAGIC J2001+439, and the same stars were then used to calibrate the NOT
    image. The uncertainty of this calibration is 0.05 mag.
    As a crosscheck, we performed a second calibration using the N-magnitudes from 25
    nearby stars from GSC2.3 and found a difference of $17 \pm 9\%$ between these two calibrations.

\begin{figure}
    \centering
    \includegraphics[width=8.0cm]{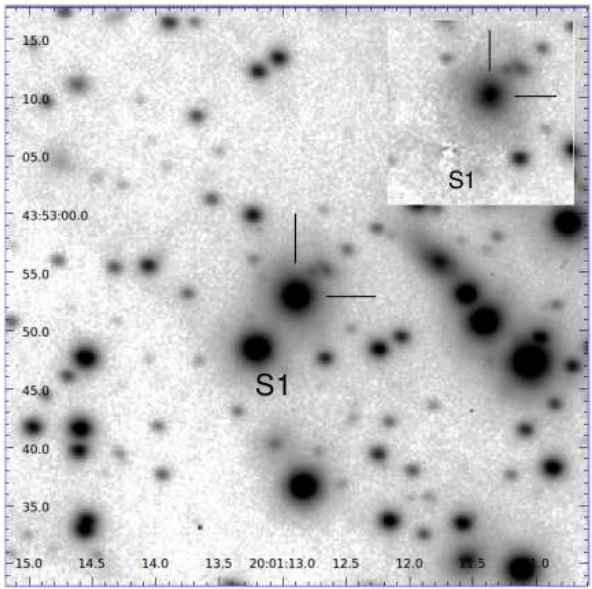}
    \caption{\label{MG4field} Portion of the NOT I-band image. 
    Field size is 50$^{\prime\prime}$$\times$50$^{\prime\prime}$. North is up, and east 
    is to the left. MAGIC J2001+439 and a nearby star (S1) are indicated. 
    The inset shows MAGIC J2001+439 and 
    S1 after subtracting a properly scaled PSF,
    showing the host galaxy more clearly.}
    \end{figure}

    To study the host galaxy, we fitted two-dimensional surface
    brightness models to the observed light distribution of
    MAGIC J2001+439. Details of this process can be found in
    \cite{1999PASP..111.1223N,2003A&A...400...95N}. In short, we first
    determined the background level around MAGIC J2001+439 and a nearby
    star S1 (Figure \ref{MG4field}) by removing the background tilt and then
    measuring empty sky regions around the targets. 
    Next, we determined the PSF from two field stars located 55 arcsec and
    92 arcsec away from MAGIC J2001+439. These fields were carefully
    selected not to
    be contaminated by foreground/background stars and roughly equal to
    MAGIC J2001+439 in peak intensity. The two field stars were close
    enough to have the same PSF. We then fitted the PSF to
    MAGIC J2001+439 and S1 and subtracted the resulting model, which
    accurately removed the star S1 (showing a good fit with the PSF model) but revealed a clear excess around
    MAGIC J2001+439. 
    The latter was then fitted with a model consisting of
    an unresolved nucleus and a host galaxy, represented by the 
    \cite{sersic1968} profile with S\'ersic index $n$ = 4 (de Vaucouleurs profile). 
    The number of free parameters in this fit was nine:
    position and magnitude of the nucleus ($x_n$,$y_n$,$m_n$), host galaxy
    position ($x_g$,$y_g$), magnitude ($m_g$), effective radius ($r_e$),
    ellipticity ($\epsilon$), and position angle (PA), which is defined counter-clockwise from the North.
    The fit was performed using pixels within 5.7 arcsec from the center of
    MAGIC J2001+439, excluding any pixels affected by overlapping targets
    and subtracting S1 prior to the fit.

\begin{figure}
    \centering
    \includegraphics[width=9.0cm]{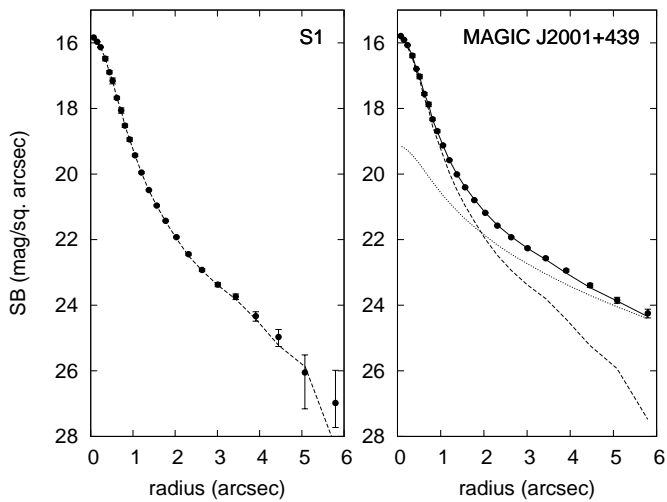}
    \caption{\label{hostradprof} Surface brightness profiles of star S1 and MAGIC J2001+439 with
    model profiles: nucleus + host galaxy (solid line), nucleus (dashed), and
    host galaxy (dotted).}
    \end{figure}

    The results of these fits are summarized in Figures \ref{MG4field} and
    \ref{hostradprof} and Table \ref{hostresults}. In addition to the best-fit values, 
    we give the errors of the fitted parameters ($\sigma_{\rm fit}$) and the calibration error
    ($\sigma_{\rm cal}$) in Table \ref{hostresults}.
    The former were estimated with $\sim$100 Monte
    Carlo simulations of the fit, which included the effects of photon
    noise, readout noise background uncertainty, and PSF variability 
    \citep[see][for details]{2003A&A...400...95N}.
    We were not able to obtain
    a perfect fit (reduced $\chi^2$ = 3.03) mainly due to the PSF mismatch
    in the core of MAGIC J2001+439 and noise in the PSF wings, both of
    which were not included in the noise model. However, our simulations
    include these effects and show that the results are not biased in any
    way due to not achieving a reduced $\chi^2$ = 1.0.


    The redshift of MAGIC J2001+439 was estimated using the observed host
    galaxy magnitude I = $17.15 \pm 0.06$ and the result by
    \cite{2005ApJ...635..173S} that the luminosities of BL Lac host
    galaxies are confined to a relatively narrow range of $M_R = -22.8 \pm
    0.5$. We used R - I = 0.7, leading to $M_I = -23.5$, the K-correction from
    \cite{1995PASP..107..945F}, and the evolution correction $E(z)=0.84*z$ to
    iteratively determine the redshift consistent with I = $17.15$ and
    $M_I = -23.5$. For the galactic extinction, we used the value
    in NED, $A_I = 0.846$, which is based on the dust reddening study by
    \cite{schlafly2011}. The redshift was estimated 1000 times with
    each time drawing $I$, $M_I$, and $A_I$ from a Gaussian distribution
    with standard deviations of 0.06, 0.5, and 0.14, respectively.
    The resulting $z$ distribution is roughly Gaussian with average
    $z = 0.18$ and standard deviation $\sigma_z = 0.04$. It should be
    noted that almost all uncertainty in this $z$ estimate arises from
    the relatively broad distribution of $M_I$.

    \begin{table}
    \caption{\label{hostresults}Host galaxy fit results. $\sigma_{\rm fit}$ gives
    the statistical error of each parameter as 
    determined by error simulations, and $\sigma_{\rm cal}$ is the calibration 
    error (see the text for further details).}
    \centering
    \begin{tabular}{llll}
    \hline\hline
    Parameter & value & $\sigma_{\rm fit}$ & $\sigma_{\rm cal}$\\
    \hline
    Nucleus magnitude $m_n$     & 16.08 & 0.02 & 0.05\\
    Host magnitude $m_g$        & 17.15 & 0.04 & 0.05\\
    Host effective radius $r_e$ & 2$\farcs$4 & 0$\farcs$4 &\\
    Host ellipticity $\epsilon$  & 0.15 & 0.03\\
    Host PA & 178$^{\circ}$ & 3$^{\circ}$ &\\
    \hline\hline
    \end{tabular}
    \end{table}

    A potential bias (systematic error) in this estimation could come
    from the assumption of the true BL Lac host galaxy luminosity
    distribution.   In this
    respect, \cite{shaw}  studied the host galaxies of 475 \emph{Fermi} BL Lacs and obtained an average host galaxy luminosity of
    $M_R = -22.5$, which is 0.3 mag fainter than the flux $M_R = -22.8 \pm 0.5$ used here \citep[retrieved from ][]{2005ApJ...635..173S}.
    If we used $M_R = -22.5$ in our calculation, we would obtain $z = 0.16 \pm 0.04$, 
    which is well within the statistical uncertainties of our measurement.
    Moreover, it should be stressed that \cite{shaw} reported that their result may be biased because they
    studied targets, which had no host detections at the time of their
    study, which means that they were probably selecting targets with fainter hosts (the
    brighter hosts were already detected by earlier authors).

    The two redshift measurements reported in section 3.3.1 and 3.3.2 ($z = 0.17 \pm 0.10$ and $z = 0.18 \pm 0.04$) 
    are compatible and consistent with the rough estimate reported by Bassani et al. (2009), and
    the lower limit reported by Shaw et al. (2013).
    The second method (using the optical measurement of the host galaxy) is more reliable because it
    uses less assumptions and yields a smaller uncertainty in the redshift value.
    We use $z = 0.18$ throughout the rest of this paper.

\section{Discussion}

\begin{figure*}[htpb]
    \centering
    \includegraphics[width=13.0cm]{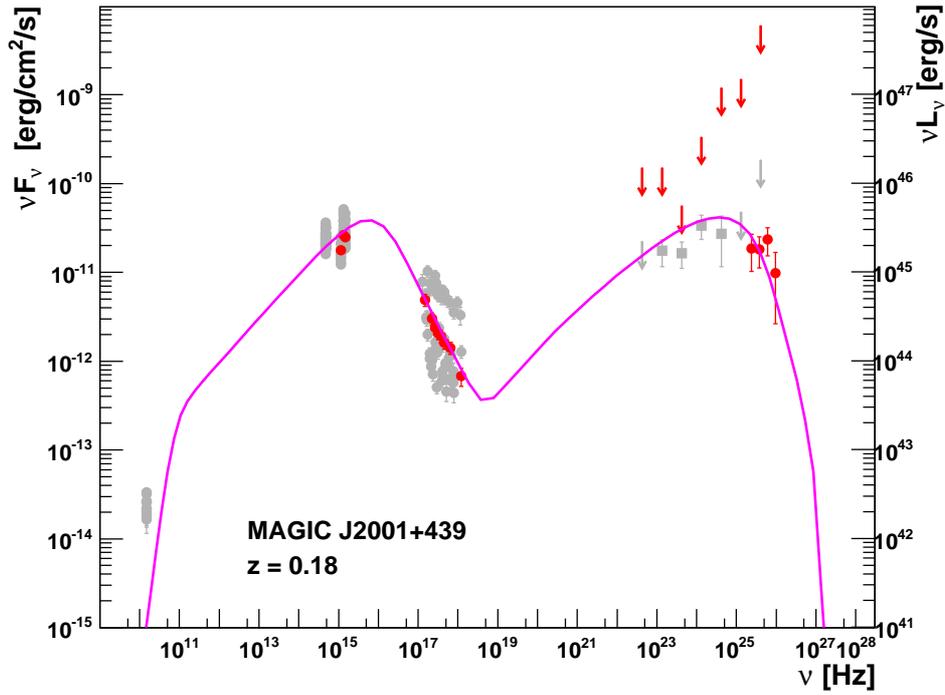}
    \caption{Broadband SED for 2010 July 16. The simultaneous data
    are depicted with red filled circles (which involve
    observations from MAGIC, \emph{Fermi}-LAT, \emph{Swift}-XRT, and \emph{Swift}-UVOT).
    The gamma-ray data points were corrected for attenuation on the EBL using a
    redshift $z$ = 0.18 and the model from \cite{franceschini2008}. 
    The red arrows show the 95\% upper limits for a one-day-averaged
    \emph{Fermi}-LAT spectrum (MJD: 55392.5 -- 55393.5), which is
    coincident with the MAGIC VHE observation from July 16.
    The gray filled squares represent the one-month-averaged
    \emph{Fermi}-LAT spectrum in 2010 July (MJD: 55378 -- 55409), and
    the gray filled circles show all the radio/optical/UV/X-ray data taken 
    during the MWL campaign in 2010, excluding July 16. The magenta solid curve
    represents the resulting intrinsic spectrum parameterized with a
    one-zone SSC model. See text for further details. }
    \label{fig13}
    \end{figure*}

    In Figure \ref{fig13}, we show the simultaneous MWL SED of MAGIC J2001+439 on 2010 July 16.
    The MAGIC data points show the deabsorbed spectrum with a redshift of $z$ = 0.18 using the EBL model of \cite{franceschini2008}.
    The deabsorbed spectrum is compatible with a simple power law with photon index $\Gamma = 2.3 \pm 0.4$.
    A one-zone SSC model, as described in the appendix of \cite{hajime2011} was used to interpret the MWL SED.
    In this model, the emission region is assumed to be spherical with radius $R$ and to be filled by a tangled magnetic field of intensity $B$ in a comoving frame.
    The emission region is in motion with a Lorentz factor of $\Gamma$ and a viewing angle of $\theta$ in the observer frame.
    The injected energy distribution of the relativistic emitting electrons is described by an unsmoothed broken power-law function\footnote[9]{The code described in
    \cite{hajime2011} can parameterize the electron energy distribution with both a smoothed and unsmoothed broken power law.},
    \begin{eqnarray}
    N (\gamma) = \left\{ \begin{array}{ll}
    K \gamma^{-n_{1}} \;\;\;\;\;\;\;\;\;\;\;\;\;\; (\gamma_{\rm{min}} < \gamma < \gamma_{\rm{bk}})\\
    K \gamma_{\rm{bk}}^{n_{2}-n_{1}} \gamma^{-n_{2}} \;\;\;\;\;\; (\gamma_{\rm{bk}} < \gamma < \gamma_{\rm{max}}),\\
    \end{array} \right.
    \end{eqnarray}
    where $K$ is the normalization factor of the electron density,
    extending from $\gamma_{\rm {min}}$ to $\gamma_{\rm {max}}$ with indices $n_{1}$ and $n_{2}$ below and above the break Lorentz factor $\gamma_{\rm {bk}}$, respectively. 
    Relativistic effects are taken into account by the Doppler factor $\delta = [\Gamma (1 - \beta \rm{cos}\theta)]^{-1}$.
    We obtained the following one-zone SSC scenario parameters: 
    $\gamma_{\rm {min}} = 1.0$, $\gamma_{\rm {bk}} = 3.9 \times 10^{4}$, $\gamma_{\rm {max}} = 6.0 \times 10^{5}$,
    $n_{1} = 2.0$, $n_{2} = 4.8$, $K = 5.2 \times 10^{3}$ cm$^{-3}$, $B = 55$ mG, $R = 20.4 \times 10^{15}$ cm and $\delta = 27$, 
    where we used the redshift $z$ = 0.18, which is the value derived from the dedicated measurement reported in section \ref{RedshiftMeasurement2}.
    The parameters $\gamma_{\rm {min}}$ and $n_{1}$ had been initially set to 1 and 2.0, respectively.
    The estimated synchrotron emission peak of MAGIC J2001+439 is located at a high frequency $\sim$ 10$^{16}$ Hz,
    which indicates that this object is 
    a typical HBL.
    The simultaneous MWL SED of MAGIC J2001+439 on 2010 July 16 can be described well by a one-zone SSC scenario.

\begin{figure*}[htpb]
    \centering
    \includegraphics[width=13.0cm]{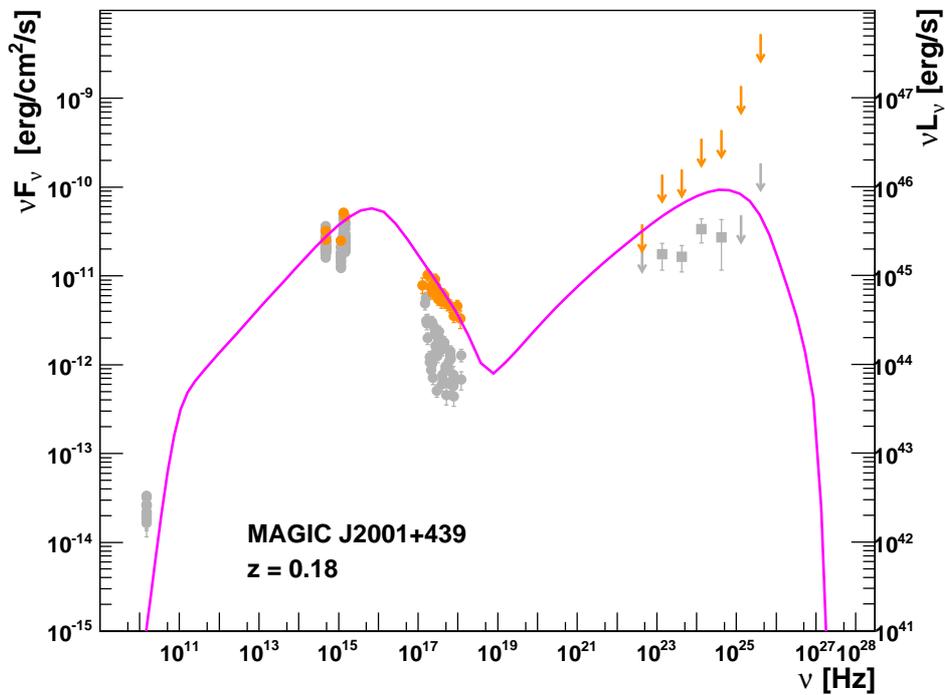}
    \caption{ Broadband SED for 2010 July 29. The simultaneous data
    are depicted with orange filled circles (which involve
    observations from \emph{Fermi}-LAT, \emph{Swift}-XRT, \emph{Swift}-UVOT and optical R-band).
    The gamma-ray data points were corrected for attenuation on the EBL using a
    redshift $z$ = 0.18 and the model from \cite{franceschini2008}. 
    The orange arrows show the 95\% upper limits for a one-day-averaged
    \emph{Fermi}-LAT spectrum (MJD: 55406 -- 55407) 
    coincident with the \emph{Swift} observations from July 29.
    The gray filled squares represent the one-month-averaged
    \emph{Fermi}-LAT spectrum in 2010 July (MJD: 55378 -- 55409), and
    the gray filled circles show all the radio/optical/UV/X-ray data taken 
    during the MWL campaign in 2010, excluding July 29.
    The magenta solid curve represents the resulting intrinsic spectrum parameterized with a
    one-zone SSC model. See text for further details.}
    \label{fig14}
    \end{figure*}

    Figure \ref{fig14} shows the SED of the X-ray flare on 2010 July 29 with the simultaneous data from \emph{Fermi}-LAT, \emph{Swift}-XRT/UVOT, and optical R-band.
    There are no VHE gamma-ray observations due to bad weather at the MAGIC site.
    The X-ray spectrum from July 29 with a photon index of $2.5 \pm 0.1$ seems to be harder than that from July 16, for which
    we obtained a photon index of $2.9 \pm 0.2$.
    The synchrotron component in the energy band between radio and X-rays shows significant variability (see Figure \ref{fig3}).
    We also find a simultaneous increase of the UV energy flux during the X-ray flare. 
    Therefore, we tried to parameterize the SED from July 29 by changing few parameters (with respect to the model used for July 16) 
    by describing the electron spectrum, 
    while keeping the environmental parameters constant of the model. 
    The obtained one-zone SSC model parameters that reproduce the observed SED data are summarized in Table 3.
    The increase and hardening of the X-ray spectrum is parameterized by an increase in the electron number density 
    ($K$ increased from $5.2 \times 10^{3}$ cm$^{-3}$ to $7.3 \times 10^{3}$ cm$^{-3}$)
    and a hardening in the slope of the electron spectrum above the break energy ($n_{2}$ changed from 4.8 to 4.3).
    This could be interpreted as an injection of fresh relativistic electrons into the emission region, which should also cause a higher
    flux in the gamma-ray bands.
    According to this theoretical scenario, both the synchrotron and the SSC bumps went up substantially during this X-ray flare.
    The observational data can only confirm the large increase in the synchrotron bump.
    However, we also note that such a change in the SED could have
    been produced by alternative scenarios, such as a change in the magnetic field strength.
    More higher quality gamma-ray data would be required to
    discriminate between different scenarios.

    \begin{figure*}[htpb]
    \centering
    \includegraphics[width=13.0cm]{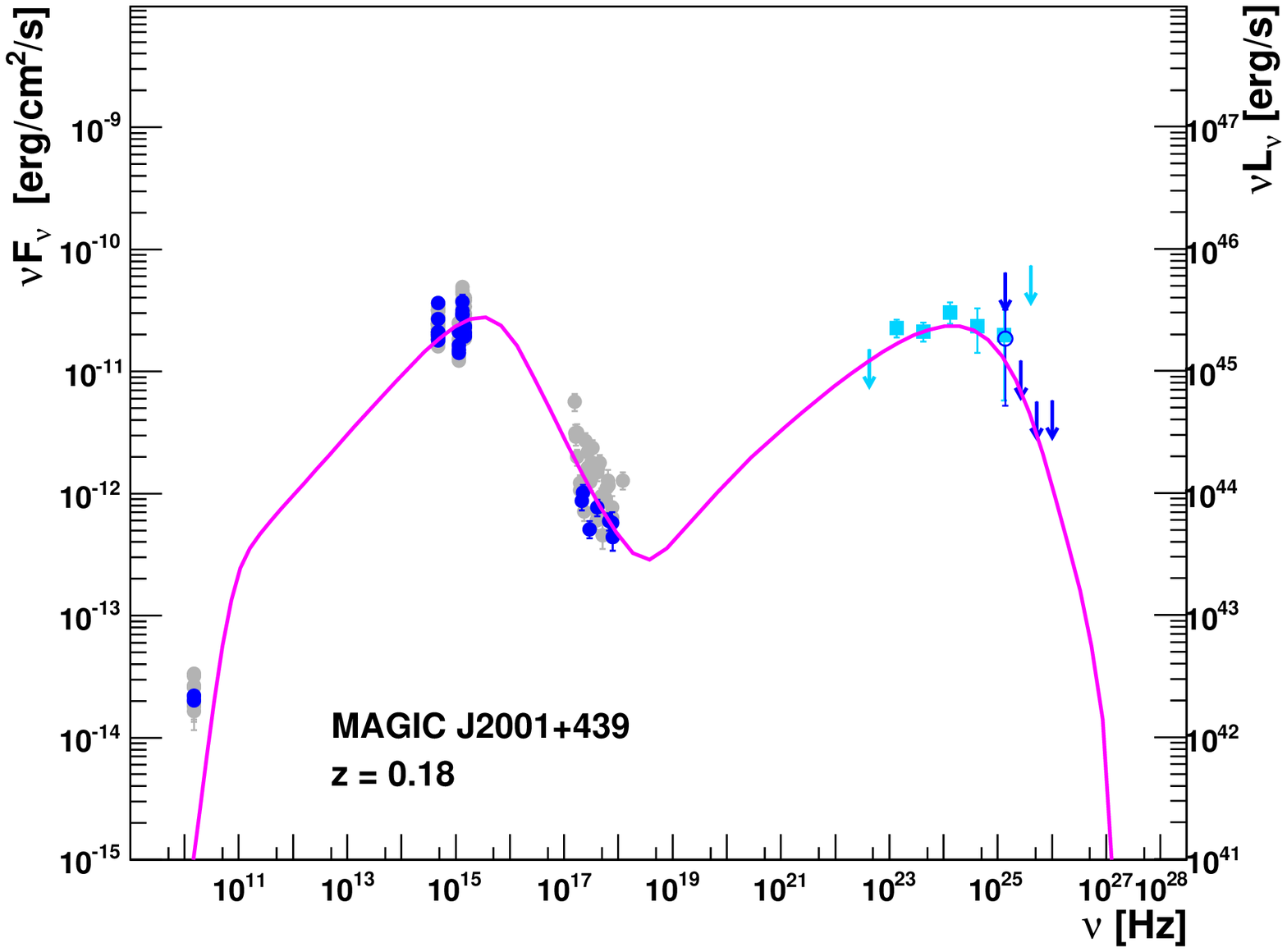}
    \caption{Contemporaneous broadband SED for the low state during the MWL campaign in 2010. 
    The blue open circle depicts the energy flux computed using all the MAGIC observations during the MWL campaign in 2010,
    after excluding the data during the VHE flare from 2010 July 16.
    This MAGIC data point contains a gamma-ray excess with a signal significance 
    \citep[calculated using Eq.17 of][]{li1983} of 1.4 $\sigma$.
    The blue arrows report the 95$\%$ confidence level energy flux upper limits derived from the MAGIC observations. 
    The cyan filled squares represent the averaged \emph{Fermi}-LAT spectrum from 
    the MWL campaign in 2010 (MJD: 55382 -- 55458).
    The cyan arrows report the \emph{Fermi}-LAT 95$\%$ confidence level energy flux
    upper limits for the energy bins with TS $<$ 4.
    The gamma-ray data points and upper limits were corrected for attenuation on the EBL using a
    redshift $z$ = 0.18 and the model from \cite{franceschini2008}. 
    The blue filled circles show the \emph{Swift}-XRT, \emph{Swift}-UVOT,
    optical, and radio data obtained during the days when MAGIC
    observed the source (with the exception of July 16).
    The gray filled circles show all the radio/optical/UV/X-ray data taken 
    during the MWL campaign in 2010, excluding July 16 and July 29. The magenta solid curve
    represents the resulting intrinsic spectrum parameterized with a
    one-zone SSC model. See text for further details.}
    \label{fig15}
    \end{figure*}

    Figure \ref{fig15} shows the contemporaneous broadband SED for the 2.5 month-long  MWL campaign.
    The figure shows both the MAGIC spectrum energy flux point with a significance (per data point)
    below 2 $\sigma$ and 95$\%$ confidence level upper limits, which were calculated assuming
    a power-law spectrum with a photon index of $\Gamma = 2.8$.
    We analyzed the 2.5 months-averaged \emph{Fermi}-LAT spectrum during the MWL campaign in 2010
    (which corresponds to the periods of the \emph{Swift} observations).
    The \emph{Fermi}-LAT spectrum between 300 MeV and 100 GeV can be characterized by
    a power-law function with a spectral index of $\Gamma_{\rm {LAT}} = 2.02 \pm 0.08$.
    The integral flux above 300 MeV is $F_{300\rm{MeV}} =$ (5.0 $\pm$ 0.6) $\times 10^{-8}$ ph cm$^{-2}$ s$^{-1}$.
    To describe the measured SED with the one-zone SSC model,
    we had to lower the value of $\gamma_{\rm{bk}}$ with respect to the one used to
    describe the SED from July 16.
    Within this theoretical scenario, this indicates that,
    the energy distribution of relativistic electrons extended to higher
    energies in July 16, when the VHE gamma-ray flare was detected with MAGIC. \\

    \begin{table*}[htpb]
    \caption{One-zone SSC model parameters for the three different SEDs reported in Figures \ref{fig13}, \ref{fig14}, and \ref{fig15}.}
    \centering
    \begin{tabular}{l c c c c c c c c c c}
    \hline\hline
    state & $\gamma_{\rm{min}}$ & $\gamma_{\rm{bk}}$ & $\gamma_{\rm{max}}$ & $n_{\rm{1}}$ & $n_{\rm{2}}$ & $K$ & $B$ & $R$ & $\delta$ & $z$ \\
     & & [$10^{4}$] & [$10^{5}$] & & & [$10^{3}$ cm$^{-3}$] & [mG] & [$10^{15}$ cm] &  & \\
    \hline
    July 16 (VHE flare detection) & 1.0 & 3.9 & 6.0 & 2.0 & 4.8 & 5.2 & 55 & 20.4 & 27 & 0.18 \\
    July 29 (X-ray flare) & 1.0 & 3.9 & 6.0 & 2.0 & 4.3 & 7.3 & 55 & 20.4 & 27 & 0.18 \\
    Typical emission during the campaign & 1.0 & 2.8 & 6.0 & 2.0 & 4.8 & 5.2 & 55 & 20.4 & 27 & 0.18 \\
    \hline\hline
    \\
    \end{tabular}
    \label{table:1}
    \end{table*}

\section{Conclusions}

    We performed a detailed study of the broadband emission of 2FGL J2001.1+4352 
    (previously named 0FGL J2001.0+4352), which had been associated to the bright radio source MG4 J200112+4352
    in \cite{bassani2009} and identified as a promising VHE emitter by the \emph{Fermi}-LAT collaboration in 2009 October.
    We characterized the radio to VHE SED and quantified the multiband
    variability and correlations over short (few days) and long (many months) timescales.

    The planned MAGIC observations led to the first VHE detection of
    this object, which we named MAGIC J2001+439.
    The multi-instrument observations showed
    variability in all the energy bands with the highest amplitude of
    variability in the X-ray and VHE bands. This source was
    significantly detected at VHE only during a 1.3 hour long MAGIC observation
    on 2010 July 16. The time-averaged VHE spectrum during this night can be
    described by a power-law function from 78 GeV to 500 GeV with a
    differential photon index $\Gamma = 2.8 \pm 0.4$ and a 
    flux normalization at 200 GeV  $f_{0}$ = $(1.9 \pm 0.4) \times 10^{-10}$
    cm$^{-2}$ s$^{-1}$ TeV$^{-1}$, which gives about 9\% the flux of
    the Crab Nebula above 200 GeV. 
    During the other nights, the VHE flux was lower than 5\% the Crab Nebula flux. 
    Besides the variability on few-day timescales, the long-term monitoring of MAGIC J2001+439 showed that
    the gamma-ray, optical, and radio emission 
    gradually decreased on few-month timescales from 2010 through 2011,
    indicating that the overall radio, optical, and gamma-ray emission is
    produced (at least a fraction of it) in a single region by the
    same population of particles. A similar positively correlated
    trend in the GeV/optical and the radio/optical bands has been
    observed on other blazars monitored over many years \citep[e.g.][]{aleksic2014a, aleksic2014b}.

    For the first time, we also determined, the redshift of this BL Lac object
    through the measurement of its host galaxy during low blazar activity.
    Because the luminosities of BL Lac host galaxies are confined to a relatively 
    narrow range \citep{2005ApJ...635..173S}, we obtained $z = 0.18 \pm 0.04$.
    Moreover, we used the \emph{Fermi}-LAT and MAGIC gamma-ray spectra to provide
    an independent redshift estimation \citep{elisa2011} by obtaining $z = 0.17 \pm 0.10$.
    The redshift values computed with these two independent methods
    are compatible within the quoted errors.
    The first method is more reliable because it uses fewer assumptions
    and yields a smaller uncertainty in the redshift value. 

    We studied the radio-to-VHE SEDs for three periods: 2010 July 16
    when the source was significantly detected at VHE with MAGIC;
    2010 July 29 when a large X-ray flux was measured (with no
    simultaneous VHE observations), and the entire dataset from the MWL
    campaign in summer 2010 with the exclusion of the observations
    from July 16 and July 29.  
    Using our redshift measurement of $z$ = 0.18,
    we described the three broadband SEDs with a one-zone SSC model.
    The model parameters that we used 
    are at the boundary of the SSC parameter distribution derived for a TeV blazar sample by \cite{tavecchio2010}.
    Within this theoretical scenario, we explain the changes
    in the broadband SEDs observed during the flaring activity in July
    16 as produced by an extension of the electron energy distribution
    towards higher energies (increase in the parameter $\gamma_{\rm{bk}}$)
    and in the SED observed during the large X-ray flare on July 29 as
    produced by an increase in the number of electrons (increase in
    the parameter $K$) and a hardening of the high-energy tail of the
    electron energy distribution (hardening of the parameter $n_2$).

    This new VHE detection adds one more BL Lac object, MAGIC
    J2001+439, to the short list of extragalactic VHE sources$^{1}$.
    Moreover, the redshift measurements we performed determined 
    that this is a relatively distant VHE BL Lac object.
    The characterization of the broadband SED with simultaneous observations 
    during various activity levels is relevant in understanding the physical properties of 
    the various blazar types and in finally moving towards AGN unification schemes.
\newline

%

\begin{acknowledgements}
   
    We would like to thank the anonymous referee for providing
    detailed and constructive remarks that helped us to improve the manuscript.
\\\\
    We would like to thank the Instituto de Astrof\'{i}sica de Canarias for the 
    excellent working conditions at the Observatorio del Roque de los Muchachos in La Palma.
    The support of the German BMBF and MPG, the Italian INFN, the Swiss National Fund SNF,
    and the Spanish MICINN is gratefully acknowledged.
    This work was also supported by the CPAN CSD2007-00042 and MultiDark CSD2009-00064 projects
    of the Spanish Consolider-Ingenio 2010 program, by grant 127740 of the Academy
    of Finland, by the DFG Cluster of Excellence "Origin and Structure of the Universe",
    by the DFG Collaborative Research Centers SFB823/C4 and SFB876/C3, and by the Polish
    MNiSzW grant 745/N-HESS-MAGIC/2010/0.      
\\\\
    The \emph{Fermi}-LAT Collaboration acknowledges support from a number of agencies and 
    institutes for both development and the operation of the LAT as well as scientific data analysis.
    These include NASA and DOE in the United States, CEA/Irfu and IN2P3/CNRS in France, ASI and 
    INFN in Italy, MEXT, KEK, and JAXA in Japan, and the K. A. Wallenberg Foundation, the Swedish
    Research Council and the National Space Board in Sweden. Additional support from INAF in Italy 
    and CNES in France for science analysis during the operations phase is also gratefully 
    acknowledged. 
\\\\
    We gratefully acknowledge the entire \emph{Swift} team, the duty scientists and science
    planners for the dedicated support, making these observations possible.
\\
    The data on this paper are based partly on
    observations made with the Nordic Optical Telescope, operated by the
    Nordic Optical Telescope Scientific Association at the Observatorio
    del Roque de los Muchachos, La Palma, Spain, of the Instituto de
    Astrofisica de Canarias. Part of the data were obtained with ALFOSC,
    which is provided by the Instituto de Astrofisica de Andalucia (IAA)
    under a joint agreement with the University of Copenhagen and NOTSA.
    The St.Petersburg University team acknowledges support from the Russian RFBR
    foundation, grant 12-02-00452. The authors thank Svetlana Jorstad for kindly providing
    the optical calibration images.
\\
    The OVRO 40~m monitoring program is supported in part by NSF grants AST-0808050 and AST-1109911 
    and NASA grants NNX08AW31G and NNX11AO43G. TH was supported by the Jenny and Antti Wihuri foundation. 
    The National Radio Astronomy Observatory is a facility of the National Science Foundation operated 
    under cooperative agreement by Associated Universities, Inc.
\\
    Special thanks to Richard Schwartz, who performed observations with the Galaxy View observatory
    and diligently reduced the data while fighting against a pancreatic cancer during the last months of his life.

\end{acknowledgements}


\bibliographystyle{plainnat}

\begin{appendix}

\section{$Swift$-XRT and $Swift$-UVOT results}

    \begin{table*}[htpb]
    \caption{Results of $Swift$-XRT observation during MWL campaign 2010; observation start time (MJD); integral flux in the energy range between 0.3 and 2 keV;
    integral flux in the energy range between 2 and 10 keV; power-law photon index, and reduced chi-square $\chi^{2}_{\nu}$ with the number of degree of freedom $n_{\rm{dof}}$.}
    \centering
    \begin{tabular}{c c c c c c c c}
    \hline\hline
    Observation date & MJD & Flux (0.3 -- 2 keV) & Flux (2 -- 10 keV) & photon index & $\chi^{2}_{\nu}$ & $n_{\rm{dof}}$  \\
     & [days] & [$10^{-12}$ erg/cm$^{2}$/s] & [$10^{-12}$ erg/cm$^{2}$/s] & (0.3 -- 10 keV) & & \\
    \hline
    July 7 & 55384.044 & 1.0 $\pm$ 0.1 & 0.9 $\pm$ 0.3 & 2.0 $\pm$ 0.6 & 0.2 & 1 \\
    July 8 & 55385.048 & 1.9 $\pm$ 0.1 & 0.6 $\pm$ 0.2 & 2.5 $\pm$ 0.5 & 0.8 & 2 \\
    July 11 & 55388.803 & 1.8 $\pm$ 0.1 & 0.6 $\pm$ 0.2 & 2.5 $\pm$ 0.6 & 1.1 & 3 \\
    July 16 & 55393.031 & 7.1 $\pm$ 0.1 & 1.2 $\pm$ 0.2 & 2.9 $\pm$ 0.2 & 1.1 & 13 \\
    July 20 & 55397.047 & 4.3 $\pm$ 0.1 & 0.9 $\pm$ 0.2 & 2.8 $\pm$ 0.3 & 1.0 & 7 \\
    July 29 & 55406.818 & 15.2 $\pm$ 0.1 & 5.8 $\pm$ 0.4 & 2.5 $\pm$ 0.1 & 0.7 & 36 \\
    Aug. 5 & 55413.024 & 7.3 $\pm$ 0.1 & 1.2 $\pm$ 0.3 & 3.0 $\pm$ 0.4 & 1.0 & 7 \\
    Aug. 10 & 55418.049 & 1.8 $\pm$ 0.1 & 0.8 $\pm$ 0.2 & 2.3 $\pm$ 0.4 & 1.3 & 3 \\
    Aug. 22 & 55430.218 & 1.3 $\pm$ 0.1 & 0.6 $\pm$ 0.3 & 2.3 $\pm$ 0.7 & 0.4 & 1 \\
    Sep. 1 & 55440.594 & 4.9 $\pm$ 0.1 & 0.7 $\pm$ 0.2 & 3.0 $\pm$ 0.4 & 0.4 & 4 \\
    Sep. 8 & 55447.090 & 1.7 $\pm$ 0.1 & 0.7 $\pm$ 0.2 & 2.4 $\pm$ 0.4 & 0.7 & 3 \\
    Sep. 12 & 55451.852 & 3.5 $\pm$ 0.1 & 1.7 $\pm$ 0.2 & 2.3 $\pm$ 0.2 & 1.5 & 10 \\
    Sep. 18 & 55457.194 & 4.5 $\pm$ 0.1 & 1.5 $\pm$ 0.3 & 2.5 $\pm$ 0.3 & 1.3 & 8 \\
    \hline\hline
    \\
    \end{tabular}
    \label{table:1}
    \end{table*}

    \begin{table*}[htpb]
    \caption{Results of $Swift$-UVOT observation during MWL campaign 2010, observation start time (MJD), and fluxes for the different three UV filters.}
    \centering
    \begin{tabular}{c c c c c}
    \hline\hline
    Observation date & MJD & UV $W1$ flux & UV $M2$ flux & UV $W2$ flux \\
     & [days] & [$10^{-11}$ erg/cm$^{2}$/s] & [$10^{-11}$ erg/cm$^{2}$/s] & [$10^{-11}$ erg/cm$^{2}$/s] \\
    \hline
    July 5 & 55382.995 & 1.8 $\pm$ 0.2 & 3.8 $\pm$ 0.5 & 2.7 $\pm$ 0.2 \\
    July 7 & 55384.046 & 1.5 $\pm$ 0.1 & 3.1 $\pm$ 0.3 & 1.9 $\pm$ 0.2 \\
    July 8 & 55385.049 & 1.4 $\pm$ 0.1 & 2.9 $\pm$ 0.3 & 2.3 $\pm$ 0.2 \\
    July 11 & 55388.808 & 1.2$\pm$ 0.1 & 2.7 $\pm$ 0.2 & 1.9 $\pm$ 0.1 \\
    July 16 & 55393.035 & 1.8 $\pm$ 0.2 & 2.6 $\pm$ 0.3 & 2.5 $\pm$ 0.2 \\
    July 20 & 55397.051 & 1.6 $\pm$ 0.1 & 3.8 $\pm$ 0.3 & 2.8 $\pm$ 0.2 \\
    July 29 & 55406.842 & 2.5 $\pm$ 0.1 & 5.2 $\pm$ 0.3 & 4.6 $\pm$ 0.2 \\
    Aug. 5 & 55413.026 & 2.5 $\pm$ 0.1 & 5.0 $\pm$ 0.4 & 3.8 $\pm$ 0.2 \\
    Aug. 10 & 55418.075 & 1.8 $\pm$ 0.1 & 3.4 $\pm$ 0.3 & 3.0 $\pm$ 0.2 \\
    Aug. 16 & 55424.063 & 2.1 $\pm$ 0.3 & 3.7 $\pm$ 0.6 & 2.1 $\pm$ 0.3 \\
    Aug. 22 & 55430.223 & 1.3 $\pm$ 0.1 & 2.6 $\pm$ 0.2 & 2.1 $\pm$ 0.1 \\
    Sep. 1 & 55440.595 & 2.5 $\pm$ 0.1 & 4.6 $\pm$ 0.3 & 4.0 $\pm$ 0.2 \\
    Sep. 8 & 55447.091 & 1.6 $\pm$ 0.1 & 2.9 $\pm$ 0.2 & 2.3 $\pm$ 0.2 \\
    Sep. 12 & 55451.851 & 1.7 $\pm$ 0.1 & 3.0 $\pm$ 0.2 & 2.5 $\pm$ 0.2 \\
    Sep. 18 & 55457.197 & 2.1 $\pm$ 0.1 & 4.3 $\pm$ 0.3 & 3.4 $\pm$ 0.2 \\
    \hline\hline
    \\
    \end{tabular}
    \label{table:1}
    \end{table*}

\end{appendix}

\end{document}